\documentclass{aa}
\usepackage{graphics,psfig,latexsym}
\usepackage{times,graphicx,epsfig}
\newcommand{\bq}{\begin{equation}}
\newcommand{\eq}{\end{equation}}
\newcommand{\bqn}{\begin{eqnarray}}
\newcommand{\eqn}{\end{eqnarray}}
\newcommand{\dd}{\mbox{\rm d}}
\begin{document}
\title
{Large scale inhomogeneity and local dynamical friction}
\author{Andreas Just\inst{1}\and
        Jorge Pe\~narrubia \inst{1,2} %$^{\ddagger}$
}

\offprints{A. Just}
\mail{just@ari.uni-heidelberg.de}
\institute{Astronomisches Rechen-Institut, M\"onchhofstra{\ss}e 12-14, 69120
Heidelberg, Germany
\and
Max-Planck-Institut f\"ur Astronomie, K\"onigsstuhl 17, 69117 Heidelberg, Germany
}

\date{Received / Accepted}

\abstract{
We investigate the effect of  a density gradient on
Chandrasekhar's dynamical friction formula based on the method of 2-body
encounters in the local approximation. We apply these generalizations to the
orbit evolution of satellite galaxies in Dark Matter haloes. We find from
the analysis
that the main influence occurs through a position-dependent maximum impact
parameter in the Coulomb logarithm, which is determined by the local
 scale-length of the density
distribution. We also show
that for eccentric orbits the explicit dependence of the Coulomb logarithm on
position yields significant differences for the standard {\it homogeneous}
force. Including the velocity dependence of
the Coulomb logarithm yields ambigous results. The orbital fits
in the first few periods are further improved, but the deviations at later
times are much larger. 
The additional
force induced by the density gradient, the {\it inhomogeneous} force, is not
antiparallel to the satellite motion and can exceed 10\% of the homogeneous
 friction force in magnitude. However,
due to the symmetry properties of the inhomogeneous force, there is a
deformation and no secular effect on the orbit
at the first order. Therefore the inhomogeneous force can be safely neglected for
the orbital evolution of satellite galaxies. 
For the homogeneous force we compare
numerical N-body calculations with semi-analytical orbits to determine
quantitatively the accuracy of the generalized formulae of the Coulomb
logarithm in the Chandresekhar approach.
 With the local scale-length as the
maximum impact parameter we find a
significant improvement of the orbital fits and a better interpretation of the
quantitative value of the Coulomb logarithm.
}

\titlerunning{Large scale inhomogeneity and local dynamical friction}

\authorrunning{A. Just \& J. Pe\~narrubia}
\maketitle
\section{Introduction\label{introd}}

Dynamical friction has a wide range of applications in stellar-dynamical systems.
It appears as the first order term in the Fokker-Planck approximation (see  e.g.
Binney \& Tremaine \cite{bin87}, hereafter BT, Eq. (8-53)) competing with the
second order diffusion terms to allow for a Maxwellian as the equilibrium
distribution. In the field of relaxation processes (dissolution of star
clusters, gravo-thermal collapse, mass segregation) the physical constraints and
dominating aspects of dynamical friction are very different to those of the
dynamical
evolution of a single massive body in a sea of light background particles.
In the latter case the friction force is independent of
the mass of the background particles, and as long as the massive body is 
far from energy equipartition
the diffusion can be neglected. Its main applications are the satellite galaxies
 in the dark matter halo (DMH) of
their parent galaxy, and super-massive black holes (SMBH) or compact star
clusters in the central bulge region of galaxies. Even in a collisional gas,
dynamical friction occurs, but corrections due to pressure forces must be
applied (Just et al. \cite{jus86}; Just \& Kegel \cite{jus90}; Ostriker
\cite{ost99} using mode analysis, and S\'anchez-Salcedo \& Brandenburg 
\cite{san99,san01} using numerical calculations).

There
are different approaches to analyse the parameter dependence and to calculate
the magnitude of the friction force, most of them based on perturbative methods
(2-body collisions: Chandrasekhar \cite{cha43}; Binney \cite{bin77}; 
Spitzer \cite{spi87}; BT; mode analysis: 
Marochnik \cite{mar68}; Kalnajs \cite{kal72}; 
Tremaine \& Weinberg \cite{tre84}; theory of linear response: Colpi et al.
\cite{col99}; statistical correlations in  the fluctuating field: Maoz
\cite{mao93}).
 Due to its complicated structure, dynamical friction shows many different
features. In 
a recent theoretical investigation Nelson \& Tremaine (\cite{nel99})
connected many physical aspects of the fluctuating gravitational field on
the basis of the linear response theory. 
For practical applications
one basic question is whether dynamical friction acting on a massive
object is dominated by local processes. 
There is a longstanding discussion on the
contribution of global modes to dynamical friction, which are known to be
 excited and enhanced by the self-gravity of the perturbation (Weinberg
\cite{wei86,wei89,wei93}; Hernquist \& Weinberg \cite{her89}).
These strong resonances may reduce the dissipation rate considerably
(Kalnajs \cite{kal72}; Tremaine \& Weinberg \cite{tre84}; 
Weinberg \cite{wei89}), but
 see also Rauch \& Tremaine (\cite{rau96}) for the opposite effect.
 For interacting galaxies with moderate mass ratios,  the global deformation
 of the main galaxy and the strong tidal forces on the perturber can enhance the
energy and momentum loss by considerable factors, which cannot be described by
the Chandrasekhar dynamical friction form
(Prugniel \& Combes \cite{pru92}; Leeuwin \& Combes \cite{lee97} and the
discussion therein). The mass of the perturber should not exceed $\sim 5\%$ of
the mass of the host object in order to neglect both the global contribution to the
dynamical friction force and the effect of deformation.

This paper is concerned with the local contribution to dynamical friction. 
 We restrict our analysis to the low mass regime with $M<0.05\,M_{\rm h}$,
where $M$ and $M_{\rm h}$ are the mass of the perturber and the host object,
respectively. For clearness we formulate most equations in terms of the orbital
motion of satellite galaxies in the DMH of the parent galaxy and give
additionally the
explicit application to a singular isothermal halo. This case will also be used
for the numerical analysis. Nevertheless the results can be used for more general
background distributions with isotropic distribution functions and also for
compact objects. As a second  application we give the corresponding equations
 for massive Black Holes moving in galactic centres.

The
basic formula of Chandrasekhar relies on an infinite homogeneous background
 with an isotropic velocity distribution function. For a Maxwellian velocity
distribution function and a massive object with $M\gg m$ it is given by
\begin{eqnarray}
\vec{\dot{v}_{\rm hom}} &=& -C_{\rm Ch} G_{\rm hom}(X) \vec{e_{\rm v_{\rm M}}}
 \quad \mbox{with} \label{Cha0}\\
  G_{\rm hom}(X)&= & \frac{\ln\Lambda}{X^2} \left[ erf(X) 
                -\frac{2X}{\sqrt{\pi}}\exp(-X^2)\right]\quad ,\nonumber\\ 
 C_{\rm Ch} &= & \frac{2\pi G^2 \rho_0 M}{\sigma^2} \quad\mbox{and}\quad
 X=\frac{|v_{\rm M}|}{\sqrt{2}\sigma} \label{Cc}
\end{eqnarray}
(BT Eqs. (7-14), (7-18)). Here $M$ and $v_{\rm M}$ are
mass and velocity of the 
massive body; $m$,
$\rho_0$, and $\sigma$ are mass, mass density, and velocity dispersion of the
ambient particles, respectively.
In this formula,  only  2-body interactions between the massive particle and
the background are included, adopting a straight-line motion in the unperturbed
potential. Essentially all
uncertainties of the relevant regime in phase space for these 
2-body encounters are hidden in the 
Coulomb logarithm $\ln\Lambda$. Different
perturbation methods lead to different interpretations of $\Lambda$:
(1) 2-body encounters: $\Lambda=b_{\rm max}/b_{\rm min}$ with maximum 
and minimum
impact parameter (BT)
(2) Mode analysis:  $\Lambda=k_{\rm max}/k_{\rm min}$ with
maximum and minimum relevant wavenumber (Kalnajs \cite{kal72})
(3) Theory of linear response: 
$\Lambda=\tau_{\rm max}/\tau_{\rm min}$ with
maximum and minimum time lag of the correlations
(Nelson \& Tremaine \cite{nel99}). 
The most important correction to Eq. \ref{Cha0} 
is due to an anisotropic velocity distribution (Binney \cite{bin77};
Statler \cite{sta88,sta91}). A numerical investigation of the anisotropy is
discussed in another paper for satellite
motion in flattened Dark Matter haloes (Pe\~narrubia \cite{pen03}; 
Pe\~narrubia et al. \cite{pen04}). In different approaches the
inhomogeneity of the background distribution is generally included (Kalnajs
 \cite{kal72}; Tremaine \& Weinberg \cite{tre84}; Colpi et al. \cite{col99}; 
Nelson \& Tremaine \cite{nel99}) but a useful scheme for applications is not 
yet available. An explicit formula for corrections due to
the local density gradient is given by Binney (\cite{bin77}), who has shown
that  the additional inhomogeneous force can be neglected for the dynamics
of galaxies in galaxy clusters. 
Maoz (\cite{mao93}) and follow-up papers ( S\'anchez-Salcedo
\cite{san99a}; Dominguez-Teneiro \& G\'omez-Flechoso \cite{dom98}) used
the fluctuation-dissipation theory with approximations similar to those
presented here to derive the energy loss
due to dynamical friction.
 Del Popolo \& Gambera (\cite{pop99}) and Del Popolo
(\cite{pop03}) compute the corrections due to a density enhancement 
spherically symmetric with respect to the massive body, which may be used to
estimate the shielding effect on the friction force. 

Numerical calculations suffered for a long time from resolution problems (e.g. 
Lin \& Tremaine \cite{lin83}; White \cite{whi83}; Bontekoe \& van Albada
\cite{bon87}; Weinberg \cite{wei93}) and, even in recent years, computations
with particle numbers below $10^5$ are frequently used 
 (Cora et al. \cite{cor97};
van den Bosch et al.
\cite{bos99}; Jiang \& Binney \cite{jia00}; Hashimoto et al. \cite{has03}). 
 Low resolution raises two major problems for the determination of orbital decay
times and quantitative numbers for the Coulomb logarithm $\ln\Lambda$. 
Due to noise, the
numerically determined Coulomb factors  $\Lambda$ vary significantly 
for the same orbits. With different seeds and evolution times for
relaxation of the unperturbed galaxy using 5,000 particles Cora et al. 
(\cite{cor97}) found a variation of fitted values for $\Lambda$ by 
almost a factor of 3. Even in the 20,000 particle runs the single orbits differ
by up to a factor of 2 in the satellite energy. Secondly, increasing the 
resolution
should lead to systematically larger values for the Coulomb logarithm, because
the range of impact parameters, where the perturbation of the motion of the
background particles is resolved, increases. This is explicitely shown in
Spinnato et al. (\cite{spi03}).
Therefore even an average over many low resolution computations underestimates
the friction force and leads to too large decay times for the satellite
galaxies.
With
the new generation of computers it is now possible to perform a larger number of
self-consistent
calculations with particle numbers above $10^6$ and the corresponding
high grid resolutions in particle-mesh codes (Vel\'azquez \& White 
\cite{vel99}; Pe\~narrubia et al. \cite{pen02};
Spinnato et al. \cite{spi03}). In these calculations one can 
 hope to suppress noise sufficiently and to also resolve
the small scale perturbations to model the full range for the Coulomb
logarithm. Bertin et al. (\cite{ber03}) also used a large
number of particles but only a low spatial resolution due to the restriction on
low order spherical harmonics. For the excitation of resonant global
modes in N-body calculations, one
order of magnitude higher particle numbers may be necessary. But this is not
our aim here. 

 The standard usage of Chandrasekhar's formula works with an individual fitting of
the constant Coulomb logarithm for each orbit. Fitting the Coulomb logarithm to
a single orbit provides in most cases an accurate
description of 
a large part of
the orbit. For this task, there is no need to improve the method.
But the Coulomb logarithm varies systematically from orbit to
orbit 
as discussed above
and the orbital fits drift away for long-living satellites
(Lin \& Tremaine \cite{lin83}; Jiang \& Binney \cite{jia00}; 
Spinnato et al.  \cite{spi03}; Pe\~narrubia et al. \cite{pen04}). 
In order to get
an improved friction formula, which can be used for a sample of orbits with
different parameters, we apply local constraints yielding a systematic variation
of the Coulomb logarithm with position (and also velocity).
The distance to the galactic centre as the maximum impact parameter for the orbital
evolution of the Magellanic Clouds was already proposed by Tremaine
(\cite{tre76}) and a position-dependent Coulomb logarithm was found numerically
by Bontekoe \& van Albada (\cite{bon87}) and Bertin et al. (\cite{ber03}). 
However the properties of a position-dependent $\ln \Lambda$
were never investigated systematically. 
 We discuss also the velocity dependence
of $\Lambda$ by excluding slow encounters.

A second effect of the local density gradient of the background distribution is
an additional 'inhomogeneous' force. Due
to the different symmetry of the density gradient, 
the net effect of the perpendicular component of 2-body encounters does not
vanish like that for the homogeneous part, but leads to an additional force not
anti-parallel to the satellite motion (see also Binney \cite{bin77}). 
We discuss the new Coulomb logarithm and the 
inhomogeneous friction term in detail.

We calculate dynamical friction using the explicit deflection of 2-body
encounters, because this approach is not based on the diffusion limit in phase
space. Therefore, in the case of a massive particle, the derivation also holds
for large angle deflections leading to a well-defined behaviour for small impact
parameters and slow encounters. Neglecting the effect of the mean tidal field on
the encounter yields the well known deflection
\begin{eqnarray}
\vec{\Delta v_{\rm M\|}} & = & \frac{2m\vec{V_{\rm 0}}}{M}
  \left[1+\frac{b^2V_{\rm 0}^4}{G^2 M^2}\right]^{-1} \nonumber\\
  & = & \frac{2Gm}{a(V_{\rm 0})}\frac{\vec{V_{\rm 0}}}{V_{\rm 0}^2} 
  \left[1+\frac{b^2}{a(V_{\rm 0})^2}\right]^{-1} \,, \label{deltavp}\\ 
\vec{\Delta v_{\rm M\bot}} & = & \frac{2m\vec{b} V_{\rm 0}^3}{GM^2} 
  \left[1+\frac{b^2V_{\rm 0}^4}{G^2 M^2}\right]^{-1} \nonumber\\
  & = & \frac{2Gm\vec{b}}{a(V_{\rm 0})^2 V_{\rm 0}} 
  \left[1+\frac{b^2}{a(V_{\rm 0})^2}\right]^{-1} \label{deltavo}\\
  && \qquad\mbox{with}\qquad
  a(V_{\rm 0})=\frac{GM}{V_{\rm 0}^2} \label{aV}
\end{eqnarray}
 (BT, Eqs. (7-10a,b)) for a given impact parameter $\vec{b}$ and relative 
 velocity
 $\vec{V_{\rm 0}}$. Here $a(V_{\rm 0})$ corresponds to the impact parameter for
a $90^{\circ}$ deflection.
From this acceleration we have to subtract the acceleration along the 
 unperturbed orbit properly in order to avoid artificial results due to
 differences in the mean field approximation. 

We start in Sect. \ref{coulog} with a detailed
discussion
of the collision rate of 2-body encounters and the structure of the Coulomb
logarithm with all parameter dependences. In Sect. \ref{dyfric} we compute
the inhomogeneous dynamical friction for an isotropic distribution function
including a reanalysis of the homogeneous friction force (corresponding to the
standard friction) for a consistent comparison of both parts. In Sects.
\ref{applic} and \ref{results} we apply the new formulae to typical satellite 
orbits in the Dark
Matter halo of the Milky Way and discuss the effect of the improved Coulomb
logarithm and of the inhomogeneous friction term. Sect. \ref{conclu} contains
a discussion of the relevance of using a position-dependent Coulomb logarithm
and of the inhomogeneous force for other applications
(super massive black holes in galactic cores) and other aspects (circularization
of the orbits).

\section{Encounter rates and the Coulomb logarithm\label{coulog}}

In this section we discuss the encounter rate and perform the integration over
impact parameters for the homogeneous and the inhomogeneous contribution to
dynamical friction. General symmetries of the force and the properties of the
Coulomb logarithm are described in detail.

\subsection{Encounter rate and mean field correction}

We use a local coordinate system $(p,q,s)$ with $s$ parallel to $\vec{V_{\rm 0}}$ and
$(p,q)$ the plane of impact parameters perpendicular to $\vec{V_{\rm 0}}$ 
(see Fig. \ref{coord}). The integration
over $s$ is substituted by the encounter rate $n(r)|\vec{V_{\rm 0}}|$  with
unperturbed background density $n(r)$ and with
$\vec{r}=\vec{r_{\rm M}}+\vec{b}$ using formally a 3-dimensional
impact parameter in vectorial form, if necessary, given by
$\vec{b}=(p,q,0)=b(\cos\phi,\sin\phi,0)$.
The rate of encounters for given relative velocity $\vec{V_{\rm 0}}$ and impact
 parameter $\vec{b}$ is then
\begin{equation}
\dd \nu(\vec{b},\vec{V_{\rm 0}}) = 
        |\vec{V_{\rm 0}}| n(\vec{r}) f(\vec{v}) \dd p \dd q \dd^3 V_{\rm 0}\,,
\end{equation}
the Dark Matter particles are counted when they are crossing the $(p,q)$-plane.
The density gradient parallel to $\vec{V_{\rm 0}}$ corresponds to an acceleration due
to the mean field because the phase space density along (unperturbed) orbits is
constant at equilibrium. The effect on the timing during the encounter can be
accounted for only when tidal fields are taken into account. Accelerations due to
 the mean field will not be included here. The density gradient perpendicular to
$\vec{V_{\rm 0}}$ changes the encounter rate directly. This effect of
the local inhomogeneity is dominant and will be included.

Since we are interested in the additional acceleration due to the grainy 
structure
of the background distribution, we have to take the difference of the
acceleration between the perturbed and the
unperturbed orbit. Particles moving along the unperturbed orbits
with impact parameter $\vec{b}$ and relative velocity $\vec{V_{\rm 0}}$ 
correspond to a 'line' of constant density
$\rho(\vec{r})f(\vec{v})\dd s \dd p \dd q \dd^3V_{\rm 0}$ along $s$,
that is, parallel to $V_{\rm 0}$. By symmetry, only the parallel component to 
$\vec{b}$ contributes. We find 
\begin{eqnarray}
\vec{\dot{v}_{\rm M0}} &=& \int G\rho(\vec{r}) f(\vec{v}) 
   \int_{-\infty}^{\infty} \frac{\vec{b}\dd s}{\sqrt{b^2+s^2}^3}
   \dd p \dd q \dd^3 V_{\rm 0} \nonumber\\
   &=&\quad \int \frac{2G\rho(\vec{r})\vec{b}}{b^2} f(\vec{v}) \dd p \dd q \dd^3 V_{\rm 0}
\end{eqnarray}
for the acceleration along the unperturbed motion.
This acceleration must be subtracted from the acceleration due to
2-body encounters yielding an effective
acceleration due to the grainy structure of the Dark Matter given by
\begin{equation}
\vec{\dot{v}_{\rm df}} = \int \vec{\Delta v_{\rm M}}\dd \nu(b,V_{\rm 0}) 
    -\vec{\dot{v}_{\rm M0}} \equiv \vec{\dot{v}_{\rm M\|}} + \vec{\dot{v}_{\rm M\bot}}\quad .
    \label{vdf}
\end{equation}
Since the correction due to the unperturbed orbits is parallel to $\vec{b}$, 
it affects only the
perpendicular component. We find from Eqs. \ref{deltavp} and \ref{deltavo}
\begin{eqnarray}
\vec{\dot{v}_{\rm M\|}} & = &  \int \frac{2G\rho(\vec{r})f(\vec{v})}{a(V_{\rm 0})}\frac{\vec{V_{\rm 0}}}{V_{\rm 0}} 
  \left[1+\frac{b^2}{a(V_{\rm 0})^2}\right]^{-1}
  \dd p \dd q \dd^3 V_{\rm 0}\,, \label{vpar0}
  \\ 
\vec{\dot{v}_{\rm M\bot}} & = & 
     \int \left( \frac{2G\rho(\vec{r})f(\vec{v})\vec{b}}{a(V_{\rm 0})^2} 
  \left[1+\frac{b^2}{a(V_{\rm 0})^2}\right]^{-1} \right.\nonumber\\
  &&\quad \left. -\frac{2G\rho(\vec{r})f(\vec{v})\vec{b}}{b^2} 
  \right) \dd p \dd q \dd^3 V_{\rm 0} \nonumber\\ 
  & = &  \int \frac{-2G\rho(\vec{r})f(\vec{v})\vec{b}}{b^2} 
  \left[1+\frac{b^2}{a(V_{\rm 0})^2}\right]^{-1}
  \dd p \dd q \dd^3 V_{\rm 0} \,. \nonumber\\ && \label{vbot0}
\end{eqnarray}
For the integration over the impact parameters we expand the local density
distribution up to the first order in the $(p,q)$-plane
\begin{equation}
\rho(r) \approx \rho_0 + \vec{b} \cdot \vec{\rho_{\rm \bot}} \quad\mbox{with}\quad
\vec{r}=\vec{r_{\rm M}}+\vec{b}\quad . \label{taylor}
\end{equation}
Here $\rho_0=\rho(\vec{r})|_{\rm r_{\rm M}}$ and
for the local density gradient we use the abbreviations
\begin{eqnarray}
\vec{\rho_r} &\equiv& \nabla \rho(\vec{r})|_{r_{\rm M}} 
        \equiv \left( \rho_x,\rho_y,\rho_z \right)
   \equiv \left( \rho_p,\rho_q,\rho_s \right) \quad \mbox{and} \quad \label{nr}\\
\vec{\rho_{\rm \bot}} &=& \vec{\rho_r} - 
        (\vec{\rho_r}\cdot\vec{e_{\rm V_{\rm 0}}})\vec{e_{\rm V_{\rm 0}}}
        = (\rho_p,\rho_q,0) \label{nbot} 
\end{eqnarray}
with coordinate systems $(x,y,z)$ oriented to the velocity of the satellite with
 $z$
axis parallel to $\vec{v_{\rm M}}$, and $(p,q,s)$ with $s$ axis parallel to
$\vec{V_{\rm 0}}$ (both are locally centred at $\vec{r_{\rm M}}$, see Fig. \ref{coord}).
The vector $\vec{\rho_{\rm \bot}}$ is the
component of the density gradient perpendicular to $\vec{V_{\rm 0}}$.
In the Taylor expansion 
an additional variation of $f(\vec{v})$ is neglected. So, for example, we do not
account for some spatial
variation of the velocity dispersion $\sigma(r)$.

\subsection{Symmetries and general properties}

There are two important features of the velocity change due to 2-body
encounters: The symmetries and the mass dependence. Both are fundamentally
different for the parallel and the perpendicular component.

$\vec{\Delta v_{\rm M\|}}$: In the limit of small angle deflections 
($b\gg a(V_{\rm 0})$) the parallel component 
is proportional to the mass $M$ leading to the well known mass segregation.
This also easily allows a separation from mean field effects. 

The parallel component is a function of $b^2$ only. Therefore, in a local
 Taylor
expansion of the distribution function, only terms even in $p$ and in $q$
contribute to the integral over the impact parameters. These are the constant
 term, and
diagonal terms of second derivatives, and so on.

The parallel component is odd in $V_{\rm 0}$. Therefore the contribution vanishes for
vanishing satellite velocity $v_{\rm M}$, if the distribution function $f(\vec{v})$ is 
symmetric.

$\vec{\Delta v_{\rm M\bot}}$: For small angles the perpendicular acceleration is
proportional to $M^2$ leading to a stronger mass dependence. On the other
hand, the acceleration falls off more steep with $V_{\rm 0}$ leading to a dominating
contribution at small $V_{\rm 0}$.

The perpendicular component is odd in $\vec{b}$, therefore terms odd in $p$ or $q$
in the Taylor expansion of the distribution function contribute to the 
integral over the impact parameters, i.e. the linear term, $3^{\rm rd}$
 derivative terms, ...

The perpendicular component is even in $V_{\rm 0}$. This has the effect of
a net acceleration also for vanishing $v_{\rm M}$ due to the asymmetric gravitational wake.

Note that these symmetries are relative to the velocity $\vec{V_{\rm 0}}$ and
 not to the velocity
$\vec{v_{\rm M}}$ of the satellite through the Dark Matter. The directions of the
accelerations relative to $v_{\rm M}$ depend on the symmetries of the distribution
function $f(\vec{v})$. The zeroth order (parallel component), that is the standard
Chandrasekhar friction, is only parallel to $v_{\rm M}$ for an isotropic 
distribution function.
The first order (the dominating term of the perpendicular component) is not
perpendicular to $v_{\rm M}$ even for an isotropic distribution function. For
vanishing $v_{\rm M}$ and isotropic velocity distribution function it is antiparallel
to the density gradient. This is equivalent to
an effective reduction of the enclosed mass ${\cal M}(r_{\rm M})$ of the Dark Matter
halo.

\subsection{Integration over the impact parameters \label{subimpact}}

Before integrating over the impact parameters we shall discuss the minimum and
maximum impact parameter interpretation.

{\bf Minimum impact parameter $b_0$:} 
For compact objects
the minimum impact parameter is irrelevant if $b_0\ll a(V_{\rm 0})$. In this
case $b_0$ can be set to zero, but for a unified formulation we will use the
Schwarzschild radius in the case of a Black hole.
Since $M \gg  m$, even in the case of large
angle deflections, the velocity change $\Delta \vec{v_{\rm M}}$ is small and can
be handled as
a perturbation (i.e. under validity of using superposition of 2-body encounters). 
For extended bodies, there is an additional 
effective $b_0$ determined by the typical radius of the
satellite. The half-mass radius $r_h$ is a good choice with a
 correction factor of
order unity, because the deflection angle is dominated by the force near the
minimum distance and is determined by the enclosed mass. 
For a Plummer sphere the half-mass radius is
given by $r_h = GM/(4.6\sigma_0^2) = 1.3r_c$ with a central velocity dispersion
 $\sigma_0$ and core radius $r_c$ (the detailed calculation of White 
(\cite{whi76})
corresponds to $b_0=1.6r_c$). An analytic calculation for
extended bodies is given by Mulder (\cite{mul83}).
For a King profile the half-mass radius is between 1/3 and 1/4.6 of
$GM/\sigma_0^2$,  again with central velocity dispersion
$\sigma_0$ of the satellite. Therefore we use for extended objects 
\begin{equation}
b_0 = r_h \approx \frac{GM}{4\sigma_0^2} \quad . \label{b0}
\end{equation}
The latter approximation, which also corresponds to the half-mass radius
of a singular isothermal sphere with
outer cut-off, will be used for the investigation of the parameter dependence of
the Coulomb logarithm.
 In case of a compact object, we use a characteristic radius for $b_0$
(the Schwarzschild radius $r_S$ for a Black Hole) and 
define the corresponding virtual velocity dispersion by
 $\sigma_0^2=GM/4b_0$ leading for a BH to
\begin{equation}
b_0 = r_S = \frac{2GM}{c^2} = \frac{GM}{4\sigma_0^2} 
        \quad\mbox{with}\quad \sigma_0^2\equiv\frac{GM}{4b_0}=\frac{c^2}{8} \label{b0BH}
\end{equation}
in order to get a consistent description.

{\bf Maximum impact parameter $b_1$:} The allowed (and effective) range of
impact parameters is bounded by the geometry and by the relevant time-scales.
The
local approximation of the distribution function 
in density (and possibly in velocity space also)
breaks down for distances
larger than the local scale-length $L$. Therefore the maximum impact parameter
should be smaller than $L$,
\begin{equation}
b < L \equiv \rho/|\nabla\rho| = \rho_0/|\vec{\rho_r}| \quad .\label{defL}
\end{equation}
By using this cut-off, we neglect more distant encounters and may underestimate
dynamical friction. 
For a singular isothermal sphere we have $L=r_{\rm M}/2$. We account for this
uncertainty by using a fitting factor $Q_2$ in the maximum impact parameter.

The straight-line approximation and the local approximation require that the
encounter time $t_{\rm coll}=2b/V_{\rm 0}$ is short compared to the local dynamical 
time-scale $t_{\rm dyn}$, which is essentially determined by the local crossing time 
of the satellite $t_{\rm dyn}=L/v_{\rm M}$
\begin{equation}
\frac{2b}{V_{\rm 0}} = t_{\rm coll} < t_{\rm dyn} = \frac{L}{v_{\rm M}} \quad .
\end{equation}
So the maximum impact parameter depends on 
$V_{\rm 0}$. Using normalized velocities $X=v_{\rm M}/(\sqrt{2}\sigma)$ and 
$W=V_{\rm 0}/(\sqrt{2}\sigma)$, it is given by
\begin{equation}
b_1 = Q_2 L \, min[1,\frac{V_{\rm 0}}{2v_{\rm M}}] 
\approx Q_2 L\sqrt{\frac{W^2}{4X^2+W^2}}\quad .
\end{equation}
In order to guarantee formally $b_1>b_0$ also
for very small $V_{\rm 0}$ we will use the analytic approximation
\begin{equation}
b_1^2 = b_0^2 + Q_2^2 L^2 \frac{W^2}{4X^2+W^2}\quad . \label{b1}
\end{equation}
Since again it is {\em a priori} not clear whether the
time-scale restriction does not exclude an important contribution of long-term
encounters, we will also carry out computations using a maximum impact
parameter given by
\begin{equation}
b_1^2 = b_0^2 + Q_1^2 L^2 \quad, \label{bm}
\end{equation}
neglecting the time-scale argument.

\subsection{Homogeneous and inhomogeneous dynamical friction\label{subinhom}}

For clarity we split the dynamical friction $\vec{\dot{v}_{\rm df}}$ (Eq. \ref{vdf})
into the part induced by the homogeneous
density term and that by the local density gradient (inhomogeneous friction)
\bq
\vec{\dot{v}_{\rm df}} \equiv \vec{\dot{v}_{\rm hom}} +\vec{\dot{v}_{\rm inh}}\quad .
\eq
By symmetry, one has that after integrating over the impact parameters, only
the parallel component $\vec{\dot{v}_{\rm M\|}}$ contributes to the homogeneous friction
 $\vec{\dot{v}_{\rm hom}}$, and the perpendicular component $\vec{\dot{v}_{\rm M\bot}}$
to the inhomogeneous friction $\vec{\dot{v}_{\rm inh}}$.
For comparison we calculate the standard case (homogeneous part, parallel
 component)
in the same way as the inhomogeneous
term. Inserting into Eq. \ref{vpar0} the Taylor
expansion of the density (Eq. \ref{taylor}) and integrating over impact
 parameters gives (the gradient term vanish by symmetry after integration over
 angle $\phi$ from $\dd p \dd q = \dd\phi b\dd b$)
\begin{eqnarray}
\vec{\dot{v}_{\rm hom}} &=& \int f(\vec{v}) \frac{2G\rho_0\vec{V_{\rm 0}}}{a(V_{\rm 0})V_{\rm 0}}
\int_{\rm b_0^2}^{b_1^2}   \left[1+\frac{b^2}{a(V_{\rm 0})^2}\right]^{-1} 
\pi \dd b^2 \dd^3 V_{\rm 0} \nonumber\\
 &=& 4\pi G\rho_0 \int  \frac{\vec{V_{\rm 0}}}{V_{\rm 0}}a(V_{\rm 0})
     \ln (\Lambda)f(\vec{v}) \dd^3 V_{\rm 0} \label{vp}
\end{eqnarray}
with
\begin{equation}
\Lambda=\sqrt{\left.\left(1+\frac{b_1^2}{a(V_{\rm 0})^2}\right)\right/
                \left(1+\frac{b_0^2}{a(V_{\rm 0})^2}\right)}\quad . \label{la}
\end{equation}
The inhomogeneous term arises from the leading contribution of the perpendicular
acceleration and gives the dominant contribution to
the acceleration perpendicular to $\vec{V_{\rm 0}}$. The integration over 
the angle $\phi$ leads to
\begin{equation}
\int_0^{2\pi} (\vec{b} \cdot \vec{\rho_{\rm \bot}}) \vec{b} \dd \phi =
  \pi b^2 \vec{\rho_{\rm \bot}} \quad .
\end{equation}
Using this result for the integration of  Eq. \ref{vbot0} over the impact
 parameters we find
\begin{eqnarray}
 \vec{\dot{v}_{\rm inh}} &=& -\int \pi G \vec{\rho_{\rm \bot}} f(\vec{v}) 
\int_{\rm b_0^2}^{b_1^2}  \left[1+\frac{b^2}{a(V_{\rm 0})^2}\right]^{-1} 
\dd b^2 \dd^3 V_{\rm 0} \nonumber\\
 &=& -2\pi G \int \vec{\rho_{\rm \bot}} a(V_{\rm 0})^2
     \ln (\Lambda)
    f(\vec{v}) \dd^3 V_{\rm 0} \quad , \label{vo}
\end{eqnarray}
where only the density gradient term of the Taylor expansion contributes.
Note that $\vec{\rho_{\rm \bot}}$, the density gradient perpendicular to 
$\vec{V_{\rm 0}}$, is a function of $\vec{V_{\rm 0}}$ leading to a more
 complicated angular integral
in velocity space. 

\subsection{The Coulomb logarithm}

\begin{figure}[t]
\centerline{
  \resizebox{0.98\hsize}{!}{\includegraphics[angle=270]{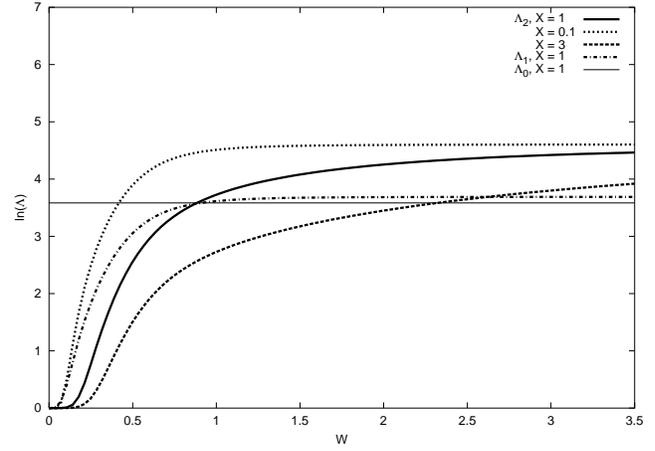}}
  }
\caption[]{
The different approximations for the Coulomb logarithm
as a function of normalized encounter velocity 
$W=V_{\rm 0}/(\sqrt{2}\sigma)$. 
The full, dotted and dashed lines show $\ln(\Lambda_2)$ (see
Eq. \ref{l2}) for different normalized satellite velocities
$X=v_{\rm M}/(\sqrt{2}\sigma)$. The satellite
parameters $q_{\rm d}=40,q_{\rm s}=5$ 
correspond to a mass of $M=5.6\cdot 10^9 M_{\rm \odot}$ at a distance of
$r_{\rm M}=40$~kpc in a Milky Way-like halo.
For comparison the Coulomb
logarithm $\ln(\Lambda_1)$ (see
Eq. \ref{l1}, dashed-dotted line) is shown. The corresponding value for a
Coulomb logarithm independent of the encounter velocity $\ln(\Lambda_0)$ 
(see Eq. \ref{l0}, thin line) is also plotted. We have chosen
$Q_0=0.8$ and $Q_2=2.5$ from the numerical experiments and the corresponding
value $Q_1=1.0$, which leads to the same friction force for $X=1$.
}
\label{figcou}
\end{figure}
For the analysis of the parameter dependence of the Coulomb logarithm, we can
reformulate Eq. \ref{la} using two scaling parameters, namely the size ratio
 $q_{\rm d}$ and the ratio of specific kinetic energies $q_{\rm s}$
defined by
\bq
q_{\rm d}=\frac{L}{b_0} \qquad q_{\rm s}=\frac{\sigma^2}{\sigma_0^2}\quad ,
\eq
instead of the dependence
 on $b_0$, $b_1$, and the mass $M$.  Since $\sigma_0$ is very large for compact
 objects (see Eq. \ref{b0BH}), we get very small values for $q_{\rm s}$, 
 which can in most cases be neglected.
With Eq. \ref{b0}, Eq. \ref{aV} becomes
\bq
a(V_{\rm 0})=\frac{GM}{V_{\rm 0}^2} = 
    \frac{2b_0}{q_{\rm s} W^2} \quad 
    .
\eq
With Eqs. \ref{b1} and \ref{la} we get the
analytic approximation
\bqn
\ln\Lambda_2 &=& \ln \sqrt{\frac{1+\frac{W^4}{4}q_{\rm s}^2
        \left(1+Q_2^2 q_{\rm d}^2\frac{W^2}{W^2+4X^2}\right)}{1+\frac{W^4}{4}q_{\rm s}^2}}
        \label{l2}\\
	&\leadsto&Q_2^2 q_{\rm d}^2q_{\rm s}^2\frac{W^6}{32X^2} 
	\quad\mbox{for}\quad
	W\leadsto 0
\eqn
and alternatively with Eq. \ref{bm} 
\bqn
\ln\Lambda_1 &=& \ln \sqrt{\frac{1+\frac{W^4}{4}q_{\rm s}^2
        \left(1+Q_1^2 q_{\rm d}^2\right)}{1+\frac{W^4}{4}q_{\rm s}^2}}
        \label{l1}\\
	&\leadsto&Q_1^2 q_{\rm d}^2q_{\rm s}^2\frac{W^4}{8} 
	\quad\mbox{for}\quad W\leadsto 0\quad .
\eqn
In order to isolate the effect of the dependence on
the encounter velocity $W$ of the Coulomb logarithm,  we also analyse
 $\ln\Lambda_0$ independent of $W$ by using the 
position-dependent maximum impact parameter $b_1=Q_0 L$ and
an estimation $a_{90}$ of the
typical impact parameter for a $90^{\circ}$ deflection. With 
$(V_0^2)_{\rm typ}=2\sigma^2 + v_{\rm M}^2$ inserted into Eq. \ref{aV} we find
\bq
a_{90} \equiv \frac{GM}{2\sigma^2 + v_{\rm M}^2} = 
        \frac{2b_0}{q_{\rm s}(1+X^2)} \quad , \label{a90}
\eq
which depends on $r_{\rm M}$ and $v_{\rm M}$, but not on $W$.  
Inserting this into Eq. \ref{la}, we find 
\bqn
\ln\Lambda_0 &=& \ln \frac{Q_0 L}{\sqrt{r_{\rm h}^2+a_{90}^2}} =
\ln \frac{Q_0 q_{\rm d}}{\sqrt{1+4q_{\rm s}^{-2}(1+X^2)^{-2}}}
        \label{l0} \\
        &\approx& \ln (Q_0 q_{\rm d})
                \qquad\quad\qquad\mbox{\rm for extended objects} \\
        &\approx& \ln \frac{Q_0L(2\sigma^2 + v_{\rm M}^2)}{GM} 
                \quad\mbox{\rm for compact objects.}    
\eqn
For the fits to the numerical
orbits we also use a globally constant value
$\langle \ln\Lambda \rangle$. 

In the case of an isothermal halo we can substitute $q_{\rm d}$ by using
${\cal M}(r_{\rm M})$,  the enclosed DMH mass, and the mass $M$ of the
 satellite, where 
\bqn
L=\frac{r_{\rm M}}{2} && {\cal M}(r_{\rm M})=4\pi\rho r_{\rm M}^3=
        \frac{2\sigma^2r_{\rm M}}{G} \\
q_{\rm m}&=&\frac{{\cal M}(r_{\rm M}) }{M}=q_{\rm s} q_{\rm d} \quad .
\eqn
 For the isothermal halo the Coulomb logarithm becomes
\bq
\ln\Lambda_2 = \ln \sqrt{\frac{1+\frac{W^4}{4}
        \left(q_{\rm s}^2+Q_2^2 q_{\rm m}^2\frac{W^2}{W^2+4X^2}\right)}
                {1+\frac{W^4}{4}q_{\rm s}^2}}\quad ,
\eq
and similar for $\ln\Lambda_1$.
In Fig. \ref{figcou} the velocity dependence of the Coulomb logarithm is
shown for different satellite velocities $X$ and typical parameters 
$q_{\rm d},q_{\rm s}$
($X=1$ corresponds to a circular orbit in an isothermal halo). We show the
velocity-dependent Coulomb logarithm $\ln(\Lambda_2)$ for 
$X=0.1;\,1.0;\,3.0$ (dotted, full, and dashed line). We have chosen
$Q_2=2.5$, which is the best fit value for typical orbits 
discussed below.
For smaller $X$
the cut-off at low relative velocities $W$ is steeper, whereas for higher $X$
 it becomes flatter. The Coulomb logarithm
$\ln(\Lambda_1)$ (dashed-dotted line) with constant $b_{\rm max}$ shows a
 shallower cut-off 
 {\bf for $W\rightarrow 0$
($\propto W^4$ for $W^2\ll 2/q_{\rm s}$ instead of 
$\propto W^6$ for $W^2\ll 2/q_{\rm s}$ and $W^2\ll 4X^2$),}
 which occurs at a lower relative velocity.
In general, the cut-off shifts the main contribution to the friction
force to higher values of $W$.
For the simpler approximation $\ln(\Lambda_1)$ we used
 $Q_1=1.0$ in order to 
get the same force for $X=1$ after integration over $W$.
Neglecting the velocity dependence by using
$\ln(\Lambda_0)$ with the best fit value $Q_0=0.9$(thin line)
leads to the corresponding constant value for comparison.
The effect of the low velocity cut-off on the friction force is discussed in
detail in Sect. \ref{dyfric}.

\section{The dynamical friction force\label{dyfric}}

For the integration over velocity space we restrict the analysis to the 
isotropic velocity
distribution functions $f(v^2)$. An anisotropy plus a
local density gradient would lead to very complicated angle integrations, which
can hardly be performed analytically. All details of the integrations in
spherical coordinates in $\dd^3V_{\rm 0}$-space are
given in the Appendix. Here we discuss the results of a Gaussian distribution
\bqn
f(v^2) &=& \frac{1}{(\sqrt{2\pi}\sigma)^3} \exp(-\frac{v^2}{2\sigma^2}) 
        \label{Gauss} 
\eqn
for the explicit integration, but the general results do not depend strongly 
on the special shape of $f(v^2)$ and it can be done in the same way for other
functions.

The homogeneous friction vector $\vec{\dot{v}_{\rm hom}}$ is parallel to
the satellite motion $\vec{v_{\rm M}}$, whereas the inhomogeneous friction lies in the
$(\vec{v_{\rm M}},\vec{\rho_r})$-plane with angle $\Psi$ 
between $\vec{v_{\rm M}}$ and $\vec{\rho_r}$. Therefore we decompose the inhomogeneous term
into the parallel and orthogonal components with respect to $\vec{v_{\rm M}}$
\bqn
\vec{\dot{v}_{\rm inh}} &\equiv& \vec{\dot{v}_{\rm par}} + \vec{\dot{v}_{\rm ort}} 
\quad\mbox{with}\quad \label{vcomp}\\ &&\qquad
\vec{\dot{v}_{\rm par}}=(\vec{\dot{v}_{\rm inh}}\cdot\vec{e_{\rm v_{\rm M}}})\vec{e_{\rm v_{\rm M}}}
\nonumber\\&&\qquad 
\vec{\dot{v}_{\rm ort}}=|\vec{\dot{v}_{\rm ort}}|\vec{e_{\rm ort}}
 \quad .\nonumber
\eqn
It is convenient to use the
normalized functions $G_{\rm hom}(X)$, $G_{\rm par}(X)$, and $G_{\rm ort}(X)$
\bqn
&&G_{\rm hom}(X) = 
  \frac{2}{\sqrt{\pi}}\int_0^\infty 
  \ln (\Lambda)   \exp(-W^2-X^2) \times 
   \nonumber \\ && \quad
\left[\cosh(2WX) - \frac{\sinh(2WX)}{2WX} \right]
        \frac{\dd W}{WX} \quad ,\label{Ghom}\\
&&G_{\rm par}(X) =
 \frac{2}{\sqrt{\pi}q_{\rm d} q_{\rm s}} 
\int_{\rm 0}^{\infty} \ln (\Lambda)  
       \exp(-W^2-X^2) \times 
   \nonumber \\ && \quad
\left[\cosh(2WX) - \frac{\sinh(2WX)}{2WX} \right]
    \frac{\dd W }{X^2 W^4} \quad , \label{Gpar}\\
&&G_{\rm ort}(X) =
 \frac{1}{\sqrt{\pi}q_{\rm d} q_{\rm s}} 
 \int_{\rm 0}^\infty \ln (\Lambda)
     \exp(-W^2-X^2) \times 
   \nonumber \\ && \quad
     \left[\sinh(2WX) \frac{4W^2 X^2+1}{2WX} - \cosh(2WX) \right]
    \frac{\dd W }{X^2 W^4} \label{Gort}
\eqn
leading to
\bqn
\vec{\dot{v}_{\rm hom}} &=& -C_{\rm Ch} G_{\rm hom}(X) \vec{e_{\rm v_{\rm M}}}
         \quad ,\label{vcha}\\
 \vec{\dot{v}_{\rm par}} &=&  
         -C_{\rm Ch} G_{\rm par}(X) \cos(\Psi)\vec{e_{\rm v_{\rm M}}} 
          \quad ,\label{vpar} \\
 \vec{\dot{v}_{\rm ort}} &=&  
         -C_{\rm Ch} G_{\rm ort}(X) \sin(\Psi)\vec{e_{\rm ort}} 
          \quad ,\label{vort}
\eqn
where the projection factors $\cos(\Psi)$ and $\sin(\Psi)$  of the density
gradient (see Eq. \ref{eort}) are separated.
The scaling factor $1/(q_{\rm d} q_{\rm s})$ ($=1/q_{\rm m}$ for an isothermal
halo) is
included in the functions $G_{\rm par}(X)$, $G_{\rm ort}(X)$. The explicit
 derivation of these functions are
given in the Appendix (see Eq. \ref{Gfun}). In a spherical halo the radial
 and
tangential components $\vec{\dot{v}_{\rm r}},\vec{\dot{v}_{\rm t}}$ of the 
inhomogeneous friction are
\bqn
 \vec{\dot{v}_{\rm r}} &=&  
         C_{\rm Ch} \left[G_{\rm ort}(X)\cos^2(i) + 
         G_{\rm par}(X)\sin^2(i)\right]\vec{e_{\rm r}} \quad ,\label{vrad} \\
 \vec{\dot{v}_{\rm t}} &=&  
         -C_{\rm Ch} \left[G_{\rm ort}(X) - G_{\rm par}(X)\right]
         \sin(i)\cos(i)\vec{e_{\rm t}} \label{vtan}
\eqn
with inclination $i$ of $\vec{\rm v_{\rm M}}$ with respect to $\vec{e_{\rm t}}$.

\subsection{Homogeneous dynamical friction\label{homfric}}

\begin{figure}[t]
\centerline{
  \resizebox{0.98\hsize}{!}{\includegraphics[angle=270]{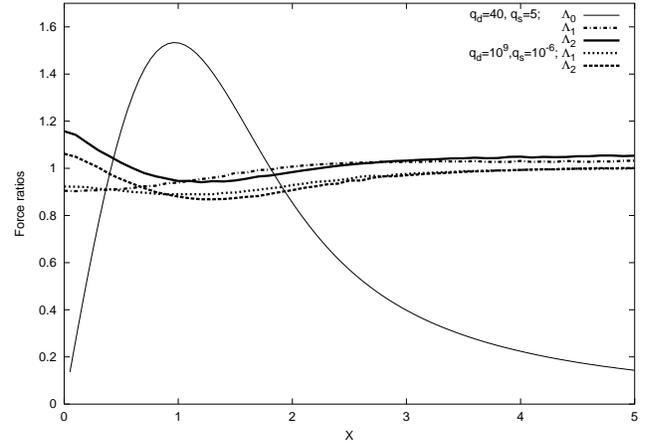}}
  }
\caption[]{
Relative homogeneous friction forces as a function of satellite velocity $X$ for the
different approximations of the Coulomb logarithm. The thin line
shows the friction force $G_{\rm hom}$ using $\ln\Lambda_0$ for orientation. 
The other lines give the relative variation 
$G_{\rm hom}(\Lambda_1)/G_{\rm hom}(\Lambda_0)$ and
$G_{\rm hom}(\Lambda_2)/G_{\rm hom}(\Lambda_0)$, respectively, for two sets of
parameters $q_{\rm d}$, $q_{\rm s}$. The first case is for a satellite galaxy,
whereas the second case corresponds to a point-like
 object (SMBH with
$M=2\,10^7 M_{\odot}$).
}
\label{figGcha}
\end{figure}

Here we shall discuss the effect of retaining the dependence of $\Lambda$ on
the encounter velocity $W$.  Neglecting $W$ in
the logarithmic term introduces an error of about 20\% (see Fig. \ref{figGcha}) 
in the friction force,
{\bf which is a function of $X$}. 
The
velocity dependence on $\ln\Lambda$ is twofold. The natural cut-off for large
angle encounters due to the reduced efficiency is included in both
approximations, $\ln\Lambda_1$ and $\ln\Lambda_2$. In $\ln\Lambda_2$
additionally the slow encounter cut-off in the maximum impact parameter (see Eq.
\ref{b1}) is included. The effect of these dependences can best be seen by
comparing with the friction force using $\ln\Lambda_0$ (Eq. \ref{l0}) with
 velocity independent
maximum impact parameter $b_1$ and an average impact parameter $a_{90}$ for the
$90^{\circ}$ deflection. $\ln\Lambda_0$ is still position and orbit dependent via $L$,
$\sigma$, and $v_{\rm M}$.
In Fig. \ref{figGcha} the force ratios using $\ln\Lambda_1$ or
$\ln\Lambda_2$ with respect to $\ln\Lambda_0$ are plotted
as a function of satellite velocity $X$ for different satellite parameters. The
typical range of $X$ for eccentric orbits is $0.5<X<2$. For an orientation the
thin line gives the normalized friction force
$G_{\rm hom}$ using $\ln\Lambda_0$. All other lines show the ratio of
$G_{\rm hom}$ using $\ln\Lambda_1$ or $\ln\Lambda_2$ with respect to that using
$\ln\Lambda_0$. We can learn three things from this plot. 
Firstly, the variation
is below 20\% for all parameter sets. Therefore for
a determination of the magnitude of the dynamical friction force, the simple
generalization $\ln\Lambda_0$ in the Chandrasekhar formula will work well. 
Secondly, the
systematic variation between apo- and peri-centre depends on the
approximation and on the parameters. This means that for a detailed
investigation of the evolution of satellite systems or of the circularization,
it may be necessary to include the explicit velocity dependence in the Coulomb
logarithm. Thirdly, there is a significant relative variation of the strength of
the forces along eccentric orbits. The ratio of the forces
at apo- and peri-centre can differ by 30\%, when using $\ln\Lambda_0$
instead of $\ln\Lambda_1$ or $\ln\Lambda_2$. Since the amount of circularization
of the orbit depends on the relative strength of the forces around apo- and
peri-centre (see Sect. \ref{orbit}), this may lead to essential
differences in the orbital evolution of satellite galaxies.
In Sects. \ref{applic} and \ref{results} we give some numerical examples to show
that the new approximations of the Coulomb 
logarithm improve the orbital fits.

\subsection{Inhomogeneous dynamical friction}

The properties of the inhomogeneous force given in Eqs. \ref{vpar} and
\ref{vort} are much more complicated. We first discuss $G_{\rm ort}(X)$, the
component orthogonal to the satellite velocity which provides the inhomogeneous
force in units of $C_{\rm Ch}$, if the density gradient is perpendicular to the
satellite motion (at peri- and apo-centre in a spherical halo). 
\begin{figure}[t]
\centerline{
  \resizebox{0.98\hsize}{!}{\includegraphics[angle=270]{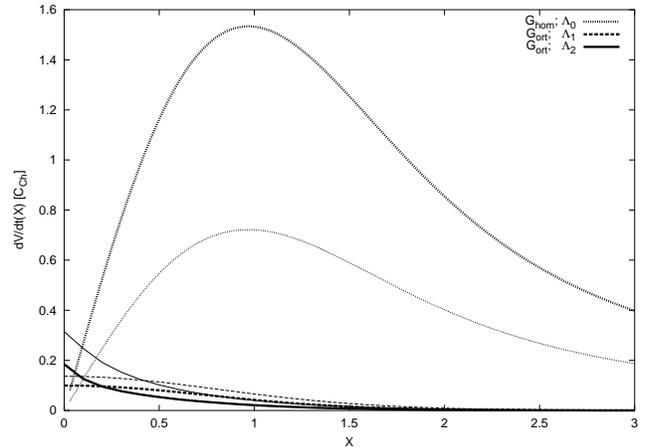}}
  }
\caption[]{
The orthogonal component of the inhomogeneous force $G_{\rm ort}(X)$ 
normalized to $C_{\rm Ch}$ is shown using
the different Coulomb logarithms: $\ln\Lambda_1$ ($Q_1=1.0$, dashed lines) and
$\ln\Lambda_2$ ($Q_2=2.5$, full lines). For
comparison the homogeneous force with velocity independent
$\ln\Lambda_0$ ($Q_0=0.9$, dotted lines) is also plotted. 
The functions are plotted for different parameters:
$q_{\rm d}=40$, $q_{\rm s}=5$ thick lines; 
$q_{\rm d}=6$, $q_{\rm s}=5$ thin lines.
}
\label{figGort}
\end{figure}
\begin{figure}[t]
\centerline{
  \resizebox{0.98\hsize}{!}{\includegraphics[angle=270]{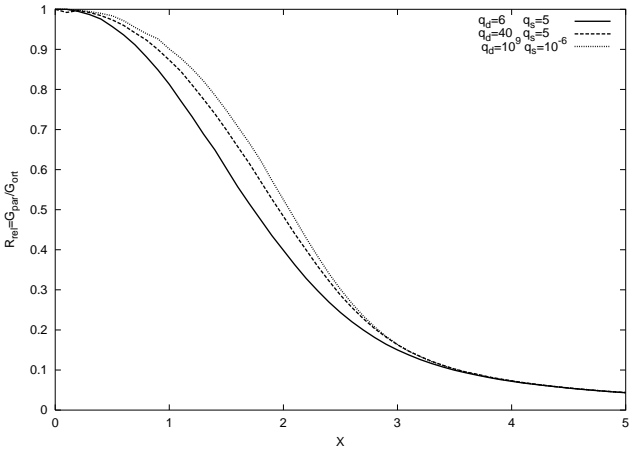}}
  }
\caption[]{
Ratio $R_{\rm rel}(X)=G_{\rm par}(X)/G_{\rm ort}(X)$ of the parallel and 
orthogonal
contribution to the inhomogeneous force (without the projection factors
$\cos\Psi$ and $\sin\Psi$) for different parameters
$q_{\rm d}$, $q_{\rm s}$.
}
\label{figGpar}
\end{figure}
In Fig. \ref{figGort} the contributions of $G_{\rm ort}$ using $\ln\Lambda_1$ and 
$\ln\Lambda_2$ are compared to $G_{\rm hom}$ with $\ln\Lambda_0$ 
(thick lines: $q_{\rm d}=40$, $q_{\rm s}=5$, 
thin lines: $q_{\rm d}=6$, $q_{\rm s}=5$). 
We have chosen
the same values $Q_0=0.9$, $Q_1=1.0$, and $Q_2=2.5$ as before. 
With $\ln\Lambda_1$ the
dependence of $G_{\rm ort}$ on $X$ is relatively flat for low satellite velocities.
In the most relevant regime $0.5<X<2$, the force can exceed 15\% of the
homogeneous force  (thin lines in Fig. \ref{figGort}). Using $\ln\Lambda_2$ leads to a steep decline for increasing
satellite velocity. For $X\approx 0$, the value is a factor of about 2
higher and for $X\approx 1$ it is a factor of 2 smaller compared to the
$\ln\Lambda_1$ case. The magnitude is up to 10\% of the homogeneous force for
 eccentricities not too large. 
The figure also shows that the
parameter dependence is quite different to the homogeneous force. The
dependence on the
relative compactness parameter $q_{\rm s}$ of $\ln\Lambda_1$ and 
$\ln\Lambda_2$ partly compensates for the reduction of $q_{\rm d}$ due to a more
extended satellite.

The component parallel to the satellite motion described by $G_{\rm par}$ is
smaller than the orthogonal contribution (see Fig. \ref{figGpar}).
The inhomogeneous force vector lies in the $(\vec{v_{\rm M}},\vec{\rho_r})$ 
plane, and the
angle $\eta$ between $\vec{\dot{v}_{\rm inh}}$ and $\vec{v_{\rm M}}$ is given by
\bq
\tan(\eta) =  \frac{\tan(\Psi)}{R_{\rm rel}(X)} \qquad \mbox{with} \qquad 
        R_{\rm rel}(X) = \frac{G_{\rm par}(X)}{G_{\rm ort}(X)}\,.
\eq
For small $X$ the force is approximately parallel to $\vec{\rho_r}$, because 
$R_{\rm rel}(0)=1$, and for large $X$ the force is approximately parallel to 
$\vec{e_{\rm ort}}$ with $R_{\rm rel}(X)\propto X^{-2}$.

In a spherical halo the
radial component points outwards along the whole orbit and the tangential
component changes sign at apo- and peri-centre
(see Eqs. \ref{vrad} and \ref{vtan}). Therefore  due to the symmetry of the
unperturbed orbits with respect to apo- and peri-centre there is no
secular effect of the inhomogeneous force. The only effect is a slight
deformation and enlargement of the orbital shape.

 Along orbits with high eccentricity, the inhomogeneous force can dominate over
the homogeneous contribution around the apo-centre, where $X\ll 1$. In this case
 there is a net
acceleration before reaching the apo-centre. The effect is an outward shift of
both the apo- and peri-centre leading to a longer duration of apo-centre passage
relative to peri-centre passage compared to the orbit without
inhomogeneous force. This may influence the evolution of highly eccentric
orbits
significantly due to an enhancement of angular momentum loss. This scenario is
very speculative, but illustrates the general properties of the inhomogeneous
force. Other effects like the bending of the gravitational wake along curved
orbits due to the time
lag may lead to a much stronger correction especially at
apo-centre. This is not included in the instantaneous and quasistationary
ansatz of the Chandrasekhar dynamical friction.

\section{Satellite motion in Dark Matter haloes\label{applic}}

In order to illustrate the effects that the different approximations for the
Coulomb logarithm induce on the satellite motion in a DMH,
we compare N-body calculations with semi-analytical computations of
typical orbits. A
comprehensive numerical study is under the way and will be presented in a
follow-up paper. We choose
parameters representative of those inferred for the Milky Way. 
We use a rigid satellite, represented by a point-mass, to avoid mass loss
and deformation effects. 
This assumption may be somewhat academic, but for the analysis of the influence
of the different approximations for the Coulomb logarithm on the orbits it is a
useful simplification. In this case 
the minimum impact parameter is
determined by the numerical resolution and not the size of the satellite.

\subsection{The N-body code}

We use {\sc Superbox} (Fellhauer et al. 2000) to evolve the
galaxy-satellite system. {\sc Superbox} is a highly efficient 
particle-mesh-code with nested and comoving grids based on a leap-frog scheme,
 and has been already implemented in
an  extensive study of satellite disruption by Klessen \& Kroupa (\cite{kle98})
 and Pe\~narrubia et al. (\cite{pen02}).

Our integration time step is $0.39$~Myr which is about $1/25$th the
dynamical time of the satellite. We have three resolution zones for the DMH,
 each
with $64^3$ grid-cells: (i) The inner grid covers the central 63~kpc in order
 to have a uniform cell-size of 2.1~kpc along the satellite orbits. 
 (ii) The middle grid
covers the whole galaxy with an extension 164~kpc giving a resolution of 5.6~kpc
per grid-cell. (iii) The outermost grid, which is common for all components,
extends to 348~kpc and contains the local universe, at a resolution of 11.6~kpc.
The satellite is modeled by an additional massive particle.

\subsection{The semi-analytical code}

We have developed a simple numerical algorithm to integrate the equations of 
motion
\bq
\vec{{\ddot r}_M}=-\nabla \Phi_g(r_{\rm M})+\vec{\dot{v}_{\rm df}}\,,
\eq
where $\Phi_g$ is the galaxy potential and $\vec{\dot{v}_{\rm df}}$ the dynamical 
friction acceleration acting on the satellite along the orbit. In this scheme,
the DMH is represented by a fixed density profile $\rho_h(r)$ 
(Eq. \ref{eqn:rho_h}), the local velocity 
dispersion is determined analytically
from the Jeans equation. The satellite is represented by a point mass with
mass $M$. 
With this kind of semi-analytical ansatz, the offset of the DMH to the
centre-of-mass ($\simeq 1$~kpc) is neglected.

\subsection{The galaxy and satellite parameters \label{galpar}}

In order to avoid the effects of the disc on the satellite evolution, we employ a
galaxy model formed purely by a DMH, with characteristics similar to those
usually assumed for the Milky Way. The density profile, therefore, can be
approximated by a quasi-isothermal sphere with total mass $M_{\rm h}$, core radius 
$\gamma$, and cut-off radius $r_{\rm cut}$ given by
\begin{equation} 
        \rho_h=\frac{M_h \alpha}{2\pi^{3/2}
        r_{\rm cut}}\frac{{\rm exp}(-r^2/r_{\rm cut}^2)}{r^2+\gamma^2},
\label{eqn:rho_h}
\end{equation}
where the normalization constant
\begin{eqnarray}
        \alpha\equiv\{1-\sqrt{\pi}\beta{\rm exp}(\beta^2)[1-{\rm erf}
                      (\beta)]\}^{-1} =  \\ \nonumber
                  1 + \sqrt{\pi}\beta + (\pi -2) \beta^2 + O(\beta^3) 
\end{eqnarray} 
having $\beta=\gamma/r_{\rm cut}$. In our
calculations we use $M_{\rm h}=1.57 \times 10^{12} \rm{M}_\odot$, 
$r_{\rm cut}=168$~kpc, and
$\gamma=3.5$~kpc. The initial distribution function in velocity space 
follows a Gaussian (Eq.~\ref{Gauss}) with $\sigma=150$~km/s and a
cut-off at
 $v=v_{\rm e}(r)=\sqrt{-2\Phi(r)}$, the escape velocity of the halo, 
approximating dynamical equilibrium.
The number of halo particles
is $N=1.4\, 10^6$ in order to get sufficient resolution. The large cut-off radius
is necessary to reach approximately an isothermal profile in the radius range of
15~kpc$<r<$60~kpc. The local scale-length drops
from $l=r/2.0$ at
$r=10$~kpc to $l=r/2.2$ at $r=55$~kpc only, and the velocity dispersion from 
$145$~km/s to $125$~km/s. The satellite is modeled by a massive particle with
mass $M=5.60\times 10^9 \rm{M}_{\rm \odot}$.
The effective minimum impact parameter when using a grid code is smaller then
the grid cell size. For the
semi-analytical analysis it seems reasonable to use for the minimum impact
parameter half the cell-size, that is $b_0=1.05$~kpc (see Spinnato et al.
 \cite{spi03} and Just \& Spurzem \cite{jus03} for a discussion).

\subsection{Numerical experiments}

\begin{table}
\begin{tabular}{lrrrrrr} \hline
{\it run} & $e$ &$r_p$ &$r_a$ &$\langle\ln\Lambda\rangle$ &$Q_0$ &$Q_2$\\ 
 &    &[kpc] &[kpc] \\  \hline
  1& 0.0 & 55.0 & 55 &2.4 &0.6 &1.9  \\
  2& 0.5 & 18.0 & 55 &2.2 &0.9 &2.5   \\
  3& 0.8 & 5.3 & 55 &1.8 &0.9 &2.8 \\ \hline
  mean& - & - & - &2.0 &0.9 &2.5 \\ \hline
\end{tabular}
\caption{The numerical experiments. The peri- and apo-galactica are    
$r_p$ and $r_a$, respectively, and $e=(r_a-r_p)/(r_a+r_p)$ is the initial
orbital eccentricity. The best fit values for the different analytical
 approximations of the
Coulomb logarithm are given in the last columns (taken from Fig. \ref{figchi}).
 The mean values are over all
 eccentricities.} 
\label{tab:numexp}
\end{table}
We started an extensive numerical study to analyse the effect of the 
improved Coulomb logarithm on orbital evolution. Here we shall discuss our first
 results concerning the quality of the
orbital fits and the different systematic behaviour with varying orbital
 eccentricity. Table \ref{tab:numexp} shows the orbital parameters,
all with apo-centre at
 $r_a=55$~kpc and peri-centre ranging down to $r_p=6.1$~kpc. In order to check
 the effect of numerical noise, we performed two additional runs for comparison. 
In detail we find: 1) Increasing the number of particles by a
factor of 8 yields an increase of the best fit Coulomb logarithm of 8\%. 
2) With a
new realization and a different orientation
of the orbit with respect to the grids, deviations of the orbits are
 below 1~kpc 
in the first 3.5~Gyr. Then the orbits run slightly out of phase, but show the
same general behaviour. 
We conclude that the basic results discussed below are not significantly disturbed
by noise.  

We start all runs at apo-centre 
 $r_0=r_{\rm M}(t=0)=55$~kpc
with a tangential velocity $v(t=0)=\epsilon v_C$ with local circular velocity
$v_C$ from the rotation curve at the satellite's
initial distance. We start with appropriate values for $\epsilon$ to reach
the desired initial eccentricities.

\section{Results \label{results}}

For a better understanding of the fitting results, we first analyse the effects
 that each analytical treatment of the Coulomb logarithm introduces in the 
orbital evolution of the satellite by only considering the dominant homogeneous
friction force. We first discuss the general orbital behaviour of the orbits.
Subsequently we analyse the differences that each dynamical friction approach
induces on the orbits. In Sect.~\ref{inhom} we consider the effect
of the inhomogeneous force on the orbit evolution.

\subsection{Orbital evolution\label{orbit}}

\begin{figure}[t]
\centerline{
  \resizebox{0.98\hsize}{!}{\includegraphics[angle=0]{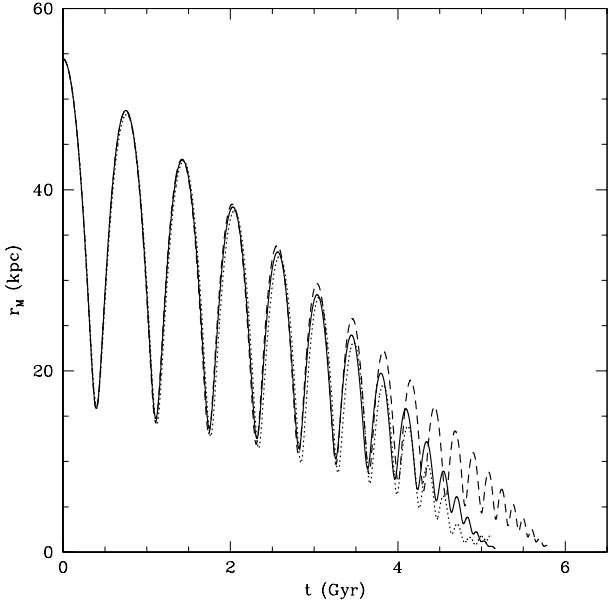}}
  }
\caption[]{Evolution of the galacto-centre distance for {\it run 2}, the live
 satellite with orbit
eccentricity $e=0.5$. The numerical data (dotted line) are compared to the 
semi-analytical
fits using a constant $\langle \ln\Lambda\rangle=2.2$ (full line), and
$\ln\Lambda_2$ with $Q_2=2.5$ (dashed line).}
\label{compr14}
\end{figure}

The orbital decay due to dynamical friction can be best observed by plotting the
galacto-centre distance evolution. For the intermediate eccentricity $e=0.5$
({\it run 2})
this is shown in Fig. \ref{compr14}. The numerical data are given as dotted
lines. 
For comparison the semi-analytical orbits with a
constant Coulomb logarithm $\langle \ln\Lambda\rangle=2.2$ (full line) and 
with $\ln\Lambda_2$ using $Q_2=2.5$ (dashed line), the respective best fit
 values, are plotted.

For the first 4 revolutions (2.5~Gyr) all fits are reasonably accurate. 
The simulations and the fits differ
systematically in the late phase, which is a well-known problem in
numerical simulations (see e.g. Bertin et al. \cite{ber03}). 
The enhanced orbital decay seems to be
connected to the large mass of the satellite. A similar effect does not occur
in numerical simulations with lower mass (e.g. Hashimoto et al. \cite{has03}).
 One
possible reason for this behaviour is the nonlinear interaction with the (live)
central region of the DMH.

\begin{figure}[t]
\centerline{
  \resizebox{0.98\hsize}{!}{\includegraphics[angle=0]{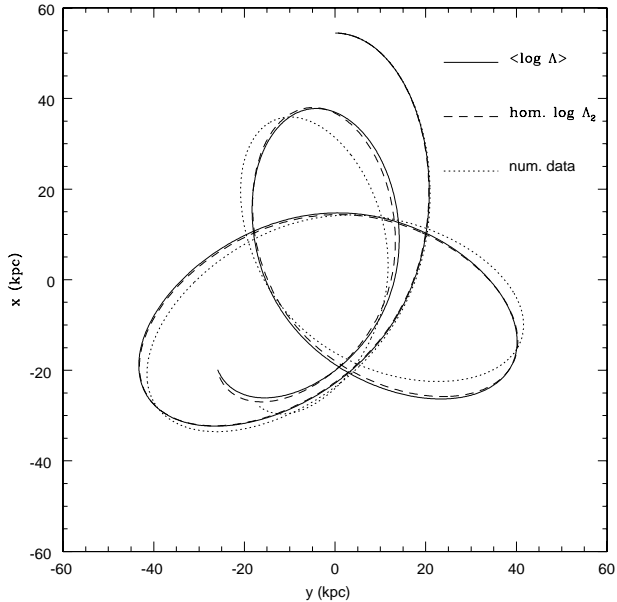}}
  }
\caption[]{Same orbits as in Fig. \ref{compr14} in the orbital plane, but
 only the first
2.5~Gyr until the fourth time apo-centre position are shown.
}
\label{proj14}
\end{figure}

Since our main goal here is to analyse the effect of the maximum impact parameter,
we restrict our investigation to the first 4 orbits. 
In Fig.~\ref{proj14} these are shown in the orbital
plane for {\it run 2}. There is a systematic delay in the
 precession rate of the N-body orbit 
compared to both semi-analytical approximations. The reason for this systematic
difference is the motion and deformation of the live halo due to the offset of
the centre-of-mass, which cannot be reproduced by an analytic approximation of
the DMH. 

\begin{figure}[t]
\centerline{
  \resizebox{0.98\hsize}{!}{\includegraphics[angle=0]{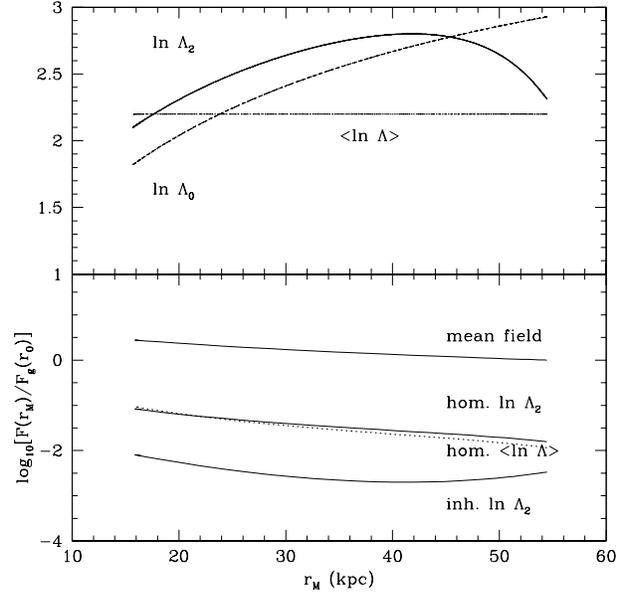}}
  }
\caption[]{The Coulomb logarithms and the dynamical friction forces along the
 first part of the orbit of {\it run 2} until
reaching the peri-centre first time.
{\bf Upper panel:} Orbital variation of the Coulomb logarithms
$\langle\ln\Lambda\rangle$ (dotted line), $\ln\Lambda_0$ (dashed line), 
and $\ln\Lambda_2(W=X)$ (full line) using the best fit values for the run.
{\bf Lower panel:} The mean field
force $G{\cal M}(r_{\rm M})/r_{\rm M}^2$, the homogeneous friction forces 
$\dot{v}_{\rm hom}$
for both Coulomb logarithms as in the upper panel, and the inhomogeneous force
$\dot{v}_{\rm inh}$ also with
$Q_2=2.5$ are plotted. All forces are normalized to the mean field force at
apo-centre.
}
\label{logvar}
\end{figure}

The Coulomb logarithm in the different approximations and therefore the
 dynamical friction force varies in a
very different way along eccentric orbits. In
Fig.~\ref{logvar} the different forces along the orbit of {\it run 2}
are shown until reaching the first peri-centre. The
upper panel shows the Coulomb logarithms $\langle\ln\Lambda\rangle$ and
$\ln\Lambda_0$. The decreasing value of $\ln\Lambda_0$ with
decreasing galacto-centric distance represents the position dependence of the
local scale-length $L$. In the case of $\ln\Lambda_2$ the value with $W=X$ is
plotted. The different shape arises from
the additional velocity dependence discussed in Sect. \ref{homfric}. The lower
panel gives the real forces along the orbit. For comparison the mean field force
$G{\cal M}(r_{\rm M})/r_{\rm M}^2$ and the inhomogeneous force are also plotted.
The most important effect of a position-dependent Coulomb logarithm is that the
relative importance of the friction force around the apo- and peri-centre is
different by up to a factor of 2 for the different approximations.

\subsection{Best fit values\label{bestfit}}

In order to quantify the accuracy of the analytical approaches and to fix the 
fitting parameter $Q$ in the Coulomb logarithm, we use the mean square
fit of the apo- and peri-centre distances by minimizing the 
parameter $\chi$ defined by 
\begin{equation}
\chi^2=\frac{1}{2k}\sum^{2k}_{\rm i=1}\left(
\left(r_{\rm i}-r_{\rm i,n}\right)^2  + \sigma_0^2\Delta t_{\rm i}^2\right)
\quad,
\label{eqn:Q}  
\end{equation}
$r_{\rm i}$ being the galacto-centric distances of the peri- and 
apo-galactica in the semi-analytical calculations.
 The subindex $n$ denotes the corresponding values of the N-body code. 
 The sum is over a given 
 number of orbits $k$. The temporal off-sets $\Delta t_{\rm i}$ are also taken
 into account weighted with the velocity dispersion at apo-centre.

\begin{figure}[t]
\centerline{
  \resizebox{0.98\hsize}{!}{\includegraphics[angle=0]{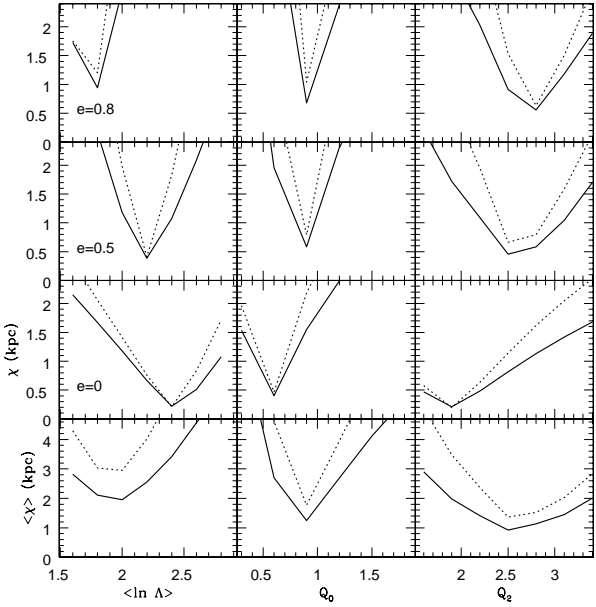}}
  }
\caption[]{Fitting factor $\chi$ as a function of the free parameters 
$\langle\ln\Lambda\rangle$, $Q_0$, and $Q_2$, respectively. The different
rows show the result for different eccentricities. The measure
of $\chi$ is done for two numbers of 
orbits, $k=3,4$ (full and dotted lines, respectively). The last row gives the
average over all eccentricities. We find well-defined minima to determine the
 best fit values listed in Table~\ref{tab:numexp}.}
\label{figchi}
\end{figure}

The resulting $\chi$-curves as a function of adopted $Q$-values are plotted
in Fig.~\ref{figchi}.
The upper rows give the values for the different eccentricities and the last row
is the average over all available eccentricities given in Tab.~\ref{tab:numexp}.
If the analytical formulae are perfectly correct, the best fit $Q$-values
should be equal to unity and should be independent of eccentricity. 
In general the $\chi$-values show well-defined minima showing that the Coulomb
logarithm is a well defined quantity. There is no big difference when using 3 or
4 orbits for the fit. We find an increasing quality of the fit
when using $\langle\ln\Lambda\rangle$, $Q_0$, and $Q_2$.
In order to understand the systematic variation of the $Q$-values, we look
 first at the constant values
$\langle\ln\Lambda\rangle=2.4,\,2.2,\,1.8$ for the different eccentricities. 
The best fit value of
$\langle\ln\Lambda\rangle$ decreases with increasing eccentricity as expected,
since the average galacto-centric distance decreases.
Looking at $Q_0$, we find that the systematic variation with eccentricity is
much smaller. From $Q_0\approx 1$, we conclude that the maximum impact
 parameter is very similar to the local
scale-length $L$. For $Q_2$ the trend changes and it seems that the radial
variation is overestimated. The large value of $Q_2\approx2.5$ shows that an
essential part of the friction force is cut off by using $\Lambda_2$ with the
time-scale argument. Slow encounters contribute essentially to the friction
force. On the
other hand, we find with $\ln\Lambda_2$ an even better fit than with
$\ln\Lambda_0$, which is already better than using the constant Coulomb
logarithm $\langle\ln\Lambda\rangle$. Due to the complicated structure of
$\ln\Lambda_2$, conclusive results for the correct fitting formula (yielding 
$Q_2\approx 1$) can be found with the full numerical investigation only.

One important goal is to improve the parameter dependence of the Coulomb
logarithm in order to use it for larger sets of orbits for a statistical
analysis of the distribution of satellite galaxies.  The variation of the best
 fit values of
$Q$ over the eccentricity range yields a smaller variation of the corresponding
Coulomb logarithms than in the case of using $\langle\ln\Lambda\rangle$.

\begin{figure}[t]
\centerline{
  \resizebox{0.98\hsize}{!}{\includegraphics[angle=0]{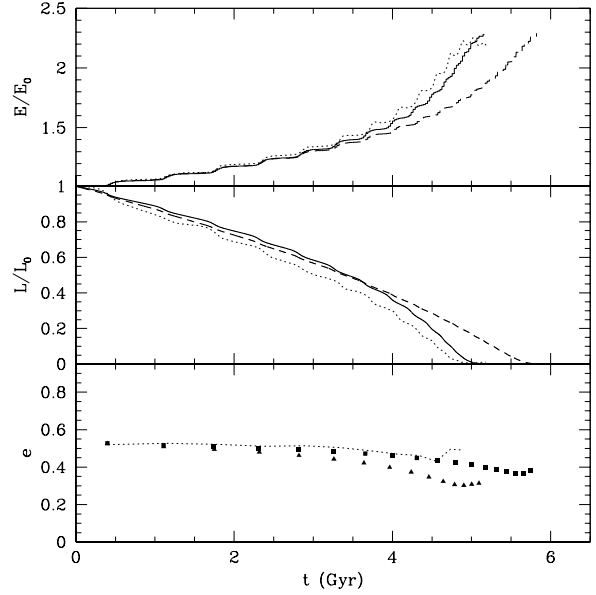}}
  }
\caption[]{Secular evolution of specific energy, angular momentum, and of the
eccentricity $e$ for the orbit of {\it run 2}. The full curve shows the case of
constant Coulomb
logarithm $\langle\ln \Lambda\rangle$, the dashed line that with $\ln
\Lambda_2$, and the
dotted line the N-body simulation. The values of $e$ are
given at peri-centre (triangles for the constant Coulomb logarithm and squares
for the varying Coulomb logarithm). The dotted line connects the numerical
values.}
\label{ele14}
\end{figure}
The evolution of specific energy $E$ and angular momentum $L$ due to drag forces
depends on the eccentricity of the orbits. Using 
$\dot{E}=v_{\rm M} \dot{v}_{\rm df}$ and
$\dot{L}=r_{\rm M} \dot{v}_{\rm df}\cos i=Lv_{\rm M}^{-1}\dot{v}_{\rm df}$
with orbit inclination $i$ we can write the ratio of
energy and angular momentum loss due to any drag force parallel to 
$\vec{v_{\rm M}}$  in the form
\bq
\frac{\dot{E}}{\dot{L}}=\frac{v_{\rm M}^2}{L} \label{edot}\,,
\eq
which holds for all points along the orbit independent of the shape of the
orbit.
The energy loss is strongly enhanced at the peri-centre, where the satellite
velocity is greater. The result for {\it run 2} with $e=0.5$ is shown in 
Fig. \ref{ele14}. The binding energy
increases in steps at the peri-centres, whereas the angular momentum decreases
quite smoothly. The fit of angular momentum is improved when using
$\ln\Lambda_2$ instead of  $\langle\ln\Lambda\rangle$. The evolution of the
 shape of the orbit, measured by the
eccentricity $e$, depends on the averages of energy and angular momentum change
over the whole revolution from
apo- to apo-centre. If the energy loss dominates, the orbit will be circularized,
and if angular momentum loss dominates, the eccentricity will increase. The
 orbit with a constant Coulomb logarithm shows a
significant circularization, which is not observed in the N-body
results (triangles compared to dotted line in the lower panel of Fig.
\ref{ele14}). The varying Coulomb logarithm essentially solves this problem
leading to constant shape of the orbit (squares in the figure). The enhanced
energy and angular momentum loss of the N-body calculation at later times will 
not  be discussed here.

\subsection{Inhomogeneous dynamical friction\label{inhom}} 

Despite the problem of determining the maximum impact parameter, the
inhomogeneity leads to the additional acceleration $\vec{\dot{v}_{\rm inh}}$,
which has, due to the
different parameter dependence and direction relative to satellite motion (see
Sect. \ref{dyfric}), a different physical effect on the orbits.

\begin{figure}[t]
\centerline{
  \resizebox{0.98\hsize}{!}{\includegraphics[angle=0]{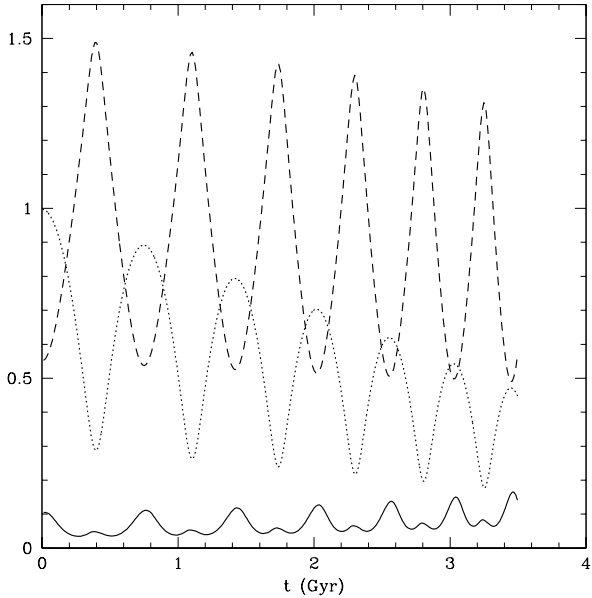}}
  }
\caption[]{Ratio of inhomogeneous to homogeneous force
$\dot{v}_{\rm inh}/\dot{v}_{\rm hom}$ 
(analytical from Eqs. \ref{vcha} - \ref{vort})
along the orbit (full line) of {\it run 2}. 
For a better 
understanding we also plot radius $r_{\rm M}/r_0$ (dotted line) and normalized
satellite velocity $X(t)$ of {\it run 2} (dashed line).}
\label{orden01}
\end{figure}

\begin{figure}[t]
\centerline{
  \resizebox{0.98\hsize}{!}{\includegraphics[angle=0]{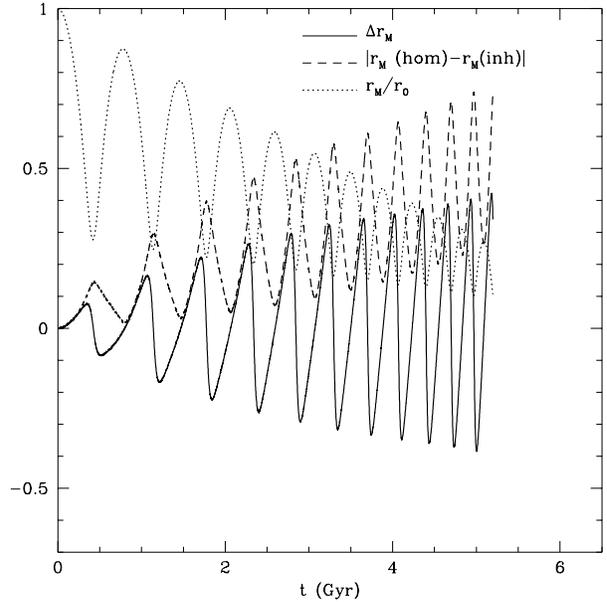}}
  }
\caption[]{Effects of including the inhomogeneous friction term
$\vec{\dot{v}_{\rm inh}}$ on the orbital evolution. Since the effects
are small we show the galacto-centric distance difference of both orbits
 $\Delta r_{\rm M}$
 (full line)
and the difference in in 3-D space 
$|\vec{r_{\rm M}}(hom)-\vec{r_{\rm M}}(inh)|$
 (dashed line) both in kpc.
The dotted line shows
the normalized position $r_{\rm M}/r_0$ of the satellite.}
\label{compr45}
\end{figure}

Since the different contributions to the force on the satellite cannot be
seperated in the numerical calculations, we discuss first the
semi-analytical forces expected along a typical orbit.
 In Fig.~\ref{orden01} we plot the ratio of the inhomogeneous and homogeneous
 force $\dot{v}_{\rm inh}/\dot{v}_{\rm hom}$ along 
 the orbit ({\it run 2} with eccentricity $e=0.5$). The maxima occur at the
 apo-galactica.
 The small bumps around the
 peri-galactica demonstrate the angle dependence of the inhomogeneous force
 being
 maximal, if the density gradient is perpendicular to the satellite motion. 
 The magnitude of the forces are shown in Fig.~\ref{logvar}. The mean value of 
 the ratio  
 slightly increases as the satellite sinks into the inner region of the galaxy,
 but it remains below 20\%.
If we compare the semi-analytical orbits with and without the inhomogeneous
force (see Fig.~\ref{compr45}), the orbit correction is smaller than 1~kpc and
dominated by a deformation of the orbit. The main effects are a slight shift of
the orbit to larger radii and a slow secular prolongation of the orbital time.
For an understanding of this small effect we can look at the decomposition
of the inhomogeneous force into the parallel and orthogonal 
 component (Eqs. \ref{vpar} and \ref{vort}).
 The parallel component 
 can be considered as a correction to the homogeneous friction force, which
 changes sign when crossing the apo- and peri-galacticon. Therefore the main
 effect is a slight outside shift of the orbit. This can also be the reason for
 the slightly increasing orbital time of the satellite. There is no 
 secular energy and momentum loss to first order. 
 The orthogonal force leads to an additional bending of the orbits.  It points
outwards and
reduces the curvature of the orbit. The decomposition into the radial and
tangential component (Eqs. \ref{vrad} and \ref{vtan}) shows that the
inhomogeneous force reduces the effect of the mean field force.
 The radial part works like a correction of the mean field by a few percent
 leading to a slightly different orbital shape with no secular part.
 
Orbit deviations of this order of magnitude cannot be
extracted from the numerical calculations,
because  even with more than $10^6$
particles the force fluctuates due to the noise in the halo particle 
distribution. From the small effect on the orbits we conclude that the
inhomogeneous force 
can safely be neglected for
the analysis of the kinematics of satellite galaxies even if the magnitude of
the inhomogeneous force is considerable. This confirms the conclusion of Binney
(\cite{bin77}) for satellite galaxies.

\section{Discussion\label{conclu}}
 
 We have recomputed the dynamical friction force on small satellite galaxies
moving in the DMH of their parent galaxies. We have extended the local approach
of Chandrasekhar by adding local constraints for the determination of the
friction force. We analysed the effect of the local density gradient of the DMH
on the standard (homogeneous) friction force and the properties of the
additional inhomogeneous force term.

\subsection{The Coulomb logarithm}

We have shown that the local scale length $L$ is an appropriate maximum impact
parameter in the Coulomb logarithm $\ln \Lambda_0$. A numerical test for an 
isothermal DMH shows
a significant improvement of orbits with different eccentricities compared to
the standard case of a
constant $\langle\ln\Lambda\rangle$. This fits to the finding of
Hashimoto at al. (\cite{has03}), who also used a position-dependent Coulomb
logarithm. The numerical results of high-resolution simulations of Spinnato et
al. (\cite{spi03}) are also in agreement with $L$ as the maximum impact
parameter. For a detailed analysis see Just \& Spurzem (\cite{jus03}). The
improved Coulomb logarithm is also in agreement with the detailed 
investigation of
Bontekoe \& van Albada (\cite{bon87}) for a n=3 polytrope with radius
 $R_{\rm out}$. They found for the
range of central distances, where the density profile is nearly exponential
(corresponding to a constant scale length $L\approx 0.1R_{\rm out}$), constant
 values for
$\ln \Lambda$, which are of the order of unity for a numerical resolution of 
$\approx 0.04R_{\rm out}$. 

In general the numerical determination of
$\ln \Lambda$ strongly depends on the numerical method and resolution. In Cora
et al. (\cite{cor97}) and other work a high level of noise and a wide
spread in the numerical determination of the Coulomb logarithm are present. An
extreme case is presented in Jiang \& Binney
(\cite{jia00}), who found $\ln \Lambda=8.5$ from orbital fits to the Sagittarius
dwarf, which corresponds to the unphysical ratio of 
$b_{\rm max}/b_{\rm min}=5,000$. They used high particle numbers but an
expansion into low order spherical harmonics with respect to the halo centre,
which strongly suppresses the local gravitational wake, and artificially enhances
the excitation of global modes by 'ghost satellites' due to the symmetry.
Cartesian meshes are best adapted to a good representation of a local 
perturbation with well-defined resolution - here global modes may be
underestimated by the lack of general symmetry.

The generalization to a velocity-dependent Coulomb logarithm by an additional
cut-off for near but slow encounters using $\ln \Lambda_2$ gives an ambiguous
result. On the one hand we find an  additional improvement of the orbit fits
for the first reolutions. 
But on the other hand in the late phase the orbital decay in the numerical
calculations is even stronger than with a constant Coulomb logarithm. The reason
may be a deformation of the central halo region due to the relative
large mass ratio $q_{\rm m}$. The linear decay when using  $\ln \Lambda_2$ is
similar to the findings of Hashimoto et al. (\cite{has03}). Additionally,
the fitting factor $Q_2\approx 2.5$ shows that by cutting
off slow encounters a significant contribution to the friction force is lost.
The rescaling by using $Q_2>1$ may change the parameter dependence essentially.
Here further investigations are necessary for conclusive results.

\subsection{Circularization}

One of the most important effects of a position-dependent $\ln \Lambda_0$ is the
 reduction of the
circularization of the orbit. 
{The variation of $\ln \Lambda$ along the orbit is a sensitive quantity for
the evolution of the orbital shape, since
the effect is significant even in the case where the dynamical friction force
is changed by about 15\% along the orbit. The circularization depends on
the systematic parameter dependence of the friction force along the orbit,
because the energy loss relative to angular momentum loss is strongly enhanced
around peri-centre due to the higher satellite velocity (see Eq. \ref{edot}). The
general effect of a reduced friction force around peri-centre is a reduction of
the energy loss leading to a lower circularization.
The effect of the position-dependent Coulomb logarithm on the circularization of
the orbits is in any case important for eccentric orbits. 
The general finding of a constant or increasing eccentricity in N-body 
simulations can be understood in this way.
 
\subsection{Inhomogeneous dynamical friction}

The inhomogeneous term
describes the effect of the asymmetry of the gravitational wake with respect to
the orbit. For the calculation of the inhomogeneous force it is essential to use the
velocity-dependent $\ln \Lambda_1$ or $\ln \Lambda_2$ in order to have a robust
 and consistent cut-off
at low impact velocities, where the dominant contribution arises (see Appendix
for a comparison with Binney (\cite{bin77}), who used a lower cut-off velocity).
The inhomogeneous force is for typical parameters of the order of
10\% relative to the homogeneous force. In highly eccentric orbits around 
apo-centre and for higher satellite masses ($\sim 10^{10}M_{\rm \odot}$) it becomes
significantly stronger. The force ratio of inhomogeneous and homogeneous terms
scales essentially with $\sim M/{\cal M}(r_{\rm M})$ (the ratio of
 satellite mass
and enclosed mass of the DMH). This is very similar to the ratio of the
orbital to the friction time-scale. 

The properties of the
inhomogeneous term are very different to the standard friction force. For an
isotropic velocity distribution it is an even function
of the satellite velocity leading to a non-vanishing force at $v_{\rm M}=0$ and to a
non-secular behaviour around apo- and peri-centre. The inhomogeneous force is
less inclined to the satellite motion than the density gradient. 
For vanishing satellite motion it points anti-parallel to the density gradient.
The parallel component gives an acceleration
during outward motion and a deceleration during inward motion. The net force
points outwards of the orbit leading to a less eccentric orbit.  

In our test orbits, the effect of the inhomogeneous force is very small leading to
an orbital offset of the order of 1~kpc over a few Gyr. Since the magnitude of
the inhomogeneous force is so small and their is no secular effect to first order,
the corrections to the friction force must be compared to other shortcomings of
the dominating homogeneous force. In highly eccentric orbits the time lag of the
perturbation leads to a stretching or squashing at peri-centre and a bending at
apo-centre of the gravitational wake due to the accelerated unperturbed motion
of the satellite. The corresponding corrections may affect the orbital evolution
much more than the inhomogeneous force.

\subsection{SMBHs in galactic centres}

For compact objects an additional velocity dependence of the Coulomb logarithm
via $a_{90}$, the typical impact parameter for a $90^0$ deflection, appears in
$\ln\Lambda_0$. 
 For an application to
the orbital decay of super-massive Black holes (SMBH) in cuspy galaxy centres, a
significant correction to the decay time-scales is required. A quantitative
 analysis of the motion of SMBHs in galactic
centres is given in Just \& Spurzem (\cite{jus03}).

\section*{Acknowledgments} 
We thank the Deutsche Forschungsgemeinschaft for supporting
JP partly through a SFB~439 grant at the University of
Heidelberg.

\section*{Appendix: Integration over velocity space}
\subsection*{Coordinate systems}
\begin{figure}[t]
\centerline{
  \resizebox{0.98\hsize}{!}{\includegraphics[angle=0]{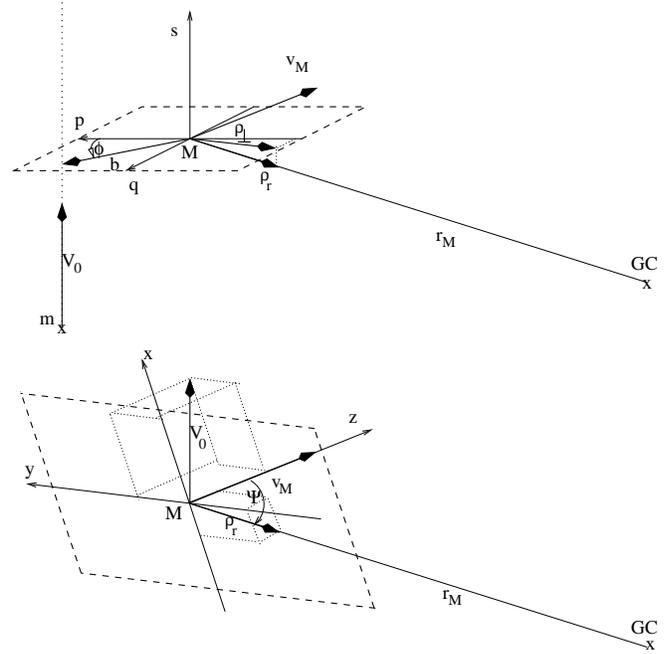}}
  }
\caption[]{
{\bf Upper panel:} Coordinate system centred at $M$ (position $\vec{r_{\rm M}}$)
oriented parallel to $\vec{V_{\rm 0}}$ for integration over impact parameter.
{\bf Lower panel:} Coordinate system centred at $M$ 
oriented parallel to $\vec{v_{\rm M}}$ for integration over velocity space $\dd^3V_{\rm 0}$.
}
\label{coord}
\end{figure}

We discuss the simplest case of an isotropic velocity distribution function 
$f(v^2)$. Since the angle integration in $\vec{v}$-space is very complicated,
for the inhomogeneous term at least, we use spherical coordinates in 
$V_{\rm 0}$-space
oriented at $v_{\rm M}$. This leads to
\bqn
\dd^3 \vec{V_{\rm 0}} &=& -V_{\rm 0}^2 \dd\varphi \dd\mu \dd V_{\rm 0}\,, \qquad \mu=\cos\theta\,,\\
\vec{V_{\rm 0}} &=& V_{\rm 0} (\sqrt{1-\mu^2}\cos\varphi,\sqrt{1-\mu^2}\sin\varphi,\mu)\,, \\
v^2 &=& |\vec{V_{\rm 0}}+\vec{v_{\rm M}}|^2 = V_{\rm 0}^2 + 2\mu V_{\rm 0}v_{\rm M} + v_{\rm M}^2\,, \\
\vec{\rho_{\rm \bot}} &=& \left( \rho_x-\cos\varphi\sqrt{1-\mu^2} 
   [\rho_x\cos\varphi\sqrt{1-\mu^2}+\right.
   \nonumber\\ &&\qquad
   \rho_y\sin\varphi\sqrt{1-\mu^2}+\rho_z\mu], 
   \nonumber\\ &&
              \rho_y-\sin\varphi\sqrt{1-\mu^2}
[\rho_x\cos\varphi\sqrt{1-\mu^2}+
   \nonumber\\ &&\qquad
\rho_y\sin\varphi\sqrt{1-\mu^2}+\rho_z\mu],
   \nonumber\\ &&
\rho_z-\mu [\rho_x\cos\varphi\sqrt{1-\mu^2}+
\nonumber\\ &&\qquad\left. 
\rho_y\sin\varphi\sqrt{1-\mu^2}+\rho_z\mu] \right)\,.  
\eqn
In the standard case the angle integrals are greatly simplified by using the 
general relation
\bqn
h(V_{\rm 0})\vec{e_{\rm V_{\rm 0}}} = \frac{\dd}{\dd \vec{V_{\rm 0}}}H(V_{\rm 0})  
     &=& -\frac{\dd}{\dd \vec{v_{\rm M}}}H(|v-v_{\rm M}|) \\
  && \qquad \mbox{with} \quad H'(V_{\rm 0})=h(V_{\rm 0})\nonumber
\eqn
and then integrating over $v$-space. But this method is not possible for the
inhomogeneous term, because the dependence of $\vec{n_{\rm \bot}}$ on $\vec{V_{\rm 0}}$
 is more
complicated. Therefore we integrate also for the homogeneous term directly over the
angles.

\subsection*{Integration over $\varphi$}

Since $v^2$ is independent of $\varphi$, we can integrate Eqs. \ref{vp} and 
\ref{vo} over $\varphi$ without
specifying the distribution function $f(v^2)$. We use
\bqn
&&\int_0^{2\pi} \vec{e_{\rm V_{\rm 0}}} \dd\varphi = 2\pi (0,0,\mu) =
         2\pi\mu\vec{e_{\rm v_{\rm M}}}\,, \\
&&\int_0^{2\pi} \vec{\rho_{\rm \bot}} \dd\varphi = 
   \pi \left( [1+\mu^2]\rho_x,[1+\mu^2]\rho_y,2[1-\mu^2]\rho_z \right) \nonumber\\
&&\,\, = \pi \frac{\rho_0}{L}\left([1+\mu^2]\sin(\Psi)\vec{e_{\rm ort}}+
        2[1-\mu^2]\cos(\Psi)\vec{e_{\rm v_{\rm M}}}\right)
\eqn
with $L=\rho_0/|\vec{\rho_r}|$ and the decomposition of the density gradient 
with respect to $\vec{v_{\rm M}}$ 
\bq
 \vec{\rho_r}=|\vec{\rho_r}|\left[\cos(\Psi)\vec{e_{\rm v_{\rm M}}}+\sin(\Psi)\vec{e_{\rm ort}}
        \right]\,,  \label{eort}
\eq
where $\vec{e_{\rm ort}}$ is the unit vector in the direction of the orthogonal 
component of $\vec{\rho_r}$  and $\Psi$ the angle
between $\vec{v_{\rm M}}$ and $\vec{\rho_r}$. We split the inhomogeneous term
in the same way into the parallel and orthogonal component 
(see Eq. \ref{vcomp}). We find for the homogeneous acceleration term and the
components of the inhomogeneous acceleration
\begin{eqnarray}
&&\vec{\dot{v}_{\rm hom}} = 4\pi G\rho_0\vec{e_{\rm v_{\rm M}}}
        \times\nonumber\\ &&\quad\int_0^\infty\int_{\rm -1}^1
        2\pi a(V_{\rm 0}) \ln(\Lambda) f(v^2) \mu\dd \mu V_{\rm 0}^2\dd V_{\rm 0} \,, \label{vhom1}\\
&&\vec{\dot{v}_{\rm par}} = \frac{-2\pi G\rho_0}{L}\cos(\Psi)\vec{e_{\rm v_{\rm M}}}
        \times\nonumber\\ &&\quad\int_0^\infty\int_{\rm -1}^1
        2\pi a(V_{\rm 0})^2 \ln(\Lambda) f(v^2) (1-\mu^2)\dd \mu V_{\rm 0}^2\dd V_{\rm 0}\,, 
         \label{vpar1}\\
&&\vec{\dot{v}_{\rm ort}} = \frac{-2\pi G\rho_0}{L}\sin(\Psi)\vec{e_{\rm ort}}
        \times\nonumber\\ &&\quad\int_0^\infty\int_{\rm -1}^1
        \pi a(V_{\rm 0})^2 \ln(\Lambda) f(v^2) (1+\mu^2)\dd \mu V_{\rm 0}^2\dd V_{\rm 0}\,. 
         \label{vort1}
\eqn

\subsection*{The integral over $\mu$}

Since $v^2$ of the Dark Matter particles is a function of $\mu$,
we must specify the distribution function before going
on. We use a Gaussian
\bqn
f(v^2) &=& \frac{1}{(\sqrt{2\pi}\sigma)^3} \exp(-\frac{v^2}{2\sigma^2}) \\
     &=& \frac{1}{(\sqrt{2\pi}\sigma)^3} \exp(-X^2-W^2)\exp(-u\mu) \\
       && \quad\mbox{with}\quad u=2WX \nonumber
\eqn
for the explicit integration, but the general results do not depend strongly 
on the
special shape of $f(v^2)$. For the homogeneous term (Eq. \ref{vhom1}) we need 
the integral
\bqn
&&
\int_{\rm -1}^1 2\pi \mu \exp(-u\mu) \dd \mu = -4\pi\left[\frac{\cosh(u)}{u} 
        - \frac{\sinh(u)}{u^2}\right]\\
    &&\qquad\qquad\approx \frac{-4\pi u}{3} \left(1 + \frac{u^2}{10} \right)
    \quad\mbox{for}\quad u\ll 1 \,, \nonumber
\eqn
and for the inhomogeneous components (Eqs. \ref{vpar1} and \ref{vort1})
the integrals
\bqn
&&\int_{\rm -1}^1 2\pi[1-\mu^2] \exp(-u\mu) \dd \mu = \nonumber\\ &&
  \qquad\qquad\quad 8\pi\left( \frac{\cosh(u)}{u^2}-
   \frac{\sinh(u)}{u^3} \right) \\
    &&\qquad\qquad\quad\approx  \frac{8\pi}{3} \left(1 + \frac{u^2}{10} \right)
    \quad\mbox{for}\quad u\ll 1 \,, \nonumber \\ 
&&\int_{\rm -1}^1 \pi[1+\mu^2] \exp(-u\mu) \dd \mu = \nonumber\\ &&
 \qquad\qquad\quad  -4\pi\left( \frac{\cosh(u)}{u^2}-
 \frac{1+u^2}{u^3}\sinh(u) \right) \\
    &&\qquad\qquad\approx  \frac{8\pi}{3} \left(1 + \frac{u^2}{5} \right)
    \quad\mbox{for}\quad u\ll 1 \,, \nonumber
\eqn
where for completeness the Taylor expansions for small velocities are also
given.

Inserting the results of the angle integration into Eq. \ref{vpar1},
 we find for the homogeneous term
\bqn
&&\vec{\dot{v}_{\rm hom}} = 
 -\frac{4\pi G^2 M\rho_0}{\sqrt{\pi}\sigma^2} \frac{\vec{v_{\rm M}}}{v_{\rm M}}
  \int_0^\infty 
  \ln \Lambda \times
       \nonumber \\ && \quad \exp(-W^2-X^2)\left(\frac{\cosh(2WX)}{WX} -
\frac{\sinh(2WX)}{2W^2X^2}\right)\dd W \nonumber \\
  &&\qquad = -C_{\rm Ch} \vec{e_{\rm v_{\rm M}}} \int_0^\infty 
      \ln (\Lambda) g_{\rm hom}(X,W) \dd W \,. \label{vch1}
\eqn
Here we used $C_{\rm Ch}$ from Eq. \ref{Cc} and the function
\bqn
g_{\rm hom}(X,W) &=& \frac{2}{\sqrt{\pi}}\exp(-W^2-X^2)\left(\frac{\cosh(2WX)}{WX} -
\right. \nonumber\\ &&\qquad\qquad\qquad \left.
                \frac{\sinh(2WX)}{2W^2X^2}\right) \label{gcha}
\eqn
($\ln\Lambda$ is a function of $V_{\rm 0}^2$, too, see Eq. \ref{la}).

For the inhomogeneous terms we get from Eq. \ref{vpar1} the parallel component
\bqn
&& \vec{\dot{v}_{\rm par}} 
 = \frac{-G^3 M^2\sqrt{\pi}\rho_r}{\sigma^4 X^2}\cos(\Psi)\vec{e_{\rm v_{\rm M}}}\times
\nonumber \\ && \quad \qquad 
\int_0^{\infty} \ln (\Lambda)  
       \exp(-W^2-X^2)\times \nonumber\\&&\qquad \qquad\qquad
    \left( \cosh(2WX) - \frac{\sinh(2WX)}{2WX} \right)
    \frac{\dd W }{W^4} \nonumber \\
  &&\quad = -C_{\rm Ch}\cos(\Psi) \vec{e_{\rm v_{\rm M}}}  
        \int_0^\infty 
      \ln (\Lambda)  g_{\rm par}(X,W)\dd W \,,
\eqn    
and from Eq. \ref{vort1} the orthogonal component
\bqn
&& \vec{\dot{v}_{\rm ort}} 
 = \frac{-G^3 M^2\sqrt{\pi}\rho_r}{2\sigma^4 X^2}\sin(\Psi)\vec{e_{\rm ort}}\times
 \nonumber \\ && \quad  \qquad
 \int_{\rm 0}^\infty \ln (\Lambda)
     \exp(-W^2-X^2)\times \nonumber\\&&\qquad\quad
     \left( \frac{4W^2 X^2+1}{2WX}\sinh(2WX) - \cosh(2WX) \right)
    \frac{\dd W }{W^4} \nonumber \\
  &&\quad = -C_{\rm Ch}\sin(\Psi) \vec{e_{\rm ort}}  
        \int_0^\infty 
      \ln (\Lambda)  g_{\rm ort}(X,W)\dd W \,,
\eqn 
where we used analogous functions
\bqn
g_{\rm par}(X,W) &=& \frac{1}{q_{\rm d} q_{\rm s}}\frac{g_{\rm hom}(X,W)}{W^3 X}  \,,\label{gpar}\\
g_{\rm ort}(X,W) &=& \frac{1}{q_{\rm d} q_{\rm s}}\frac{2h(X,W)-g_{\rm hom}(X,W)}{2W^3 X}
\quad\mbox{with}  \label{gort}\\
h(X,W) &=& \frac{2}{\sqrt{\pi}}\exp(-W^2-X^2)\sinh(2WX) \,.\nonumber
\eqn

Now we can discuss the effect of the different approximations for the Coulomb
logarithm (Eqs. \ref{l2}, \ref{l1}, \ref{l0}). For the homogeneous force
(arising from $\ln(\Lambda)g_{\rm hom}(X,W)$) the effect of the different Coulomb
logarithms on the contributions as a function of $W$ is shown in Fig.
\ref{justfiggcha}.
\begin{figure}[t]
\centerline{
  \resizebox{0.98\hsize}{!}{\includegraphics[angle=270]{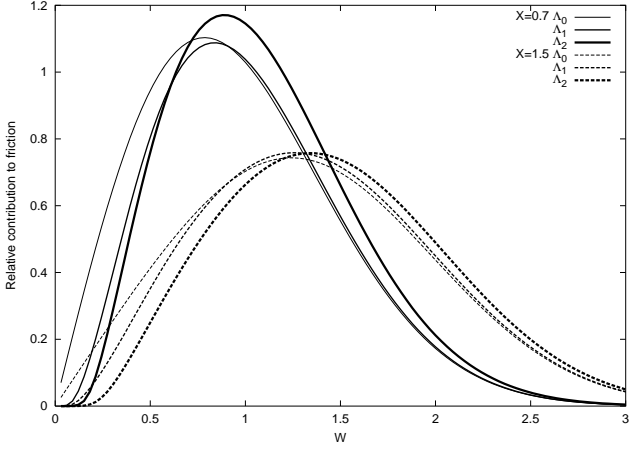}}
  }
\caption[]{
Here we compare the contribution to the homogeneous dynamical friction as a
function of encounter velocity $W$ for different satellite velocities
$X=0.7, 1.5$ using the different approximations for the Coulomb logarithm.
We chose values $Q_0=0.9$ (corresponding to the standard dynamical friction;
thin lines),
$Q_1=1.0$ (normal lines), and $Q_2=2.5$ (thick lines) leading all to the similar
homogeneous friction
for $X=1$. The system parameters are $q_{\rm d}=40$ and $q_{\rm s}=5$.
}
\label{justfiggcha}
\end{figure}
The main effect of a varying Coulomb
logarithm (with $Q_1$ or $Q_2$) is a shift of the dominant contribution to
higher encounter velocities. But the overall deformation remains small and the
dependence on the satellite motion is weak. Therefore one can use for the
homogeneous friction force the approximation of a velocity-independent Coulomb
logarithm $\ln\Lambda_0$. The relative variation with $X$ and dependence on the
system parameters $q_{\rm d},q_{\rm s}$ is shown in Fig.
\ref{figGcha}.

\begin{figure}[t]
\centerline{
  \resizebox{0.98\hsize}{!}{\includegraphics[angle=270]{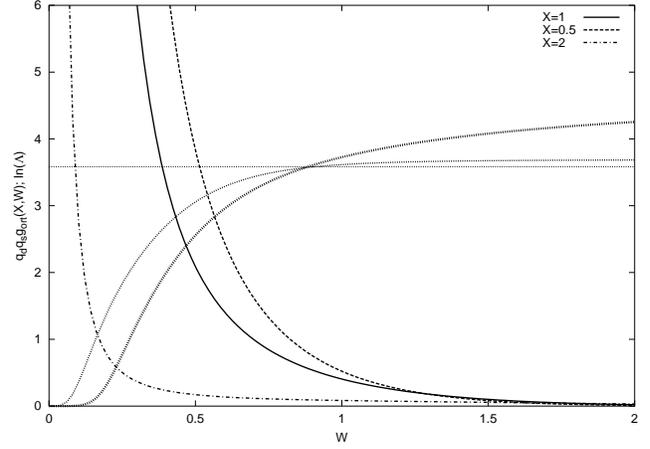}}
  }
\caption[]{
Here the function $q_{\rm d} q_{\rm s}g_{\rm ort}(X,W)$ (see Eq. \ref{gort})
for the orthogonal component of the 
inhomogeneous dynamical friction as a
function of encounter velocity $W$ for different satellite velocities
$X=0.5,1.0,2.0$ is shown. It must be multiplied by the Coulomb logarithm, before
 integrating over
$W$. The effect of using different approximations for the 
Coulomb logarithm is essential, because the dominating contribution comes from
small $W$, where $g_{\rm ort}(X,W)$ has a singularity. For imagination the
Coulomb logarithms with $q_{\rm d}=40$, $q_{\rm s}=5$, and $X=1$ from 
Fig. \ref{figcou} are plotted
again (with $Q_0=0.9$, $Q_1=1.0$, $Q_2=2.5$ thin, normal, thick dotted line,
respectively).
}
\label{justfiggort1}
\end{figure}
\begin{figure}[t]
\centerline{
  \resizebox{0.98\hsize}{!}{\includegraphics[angle=270]{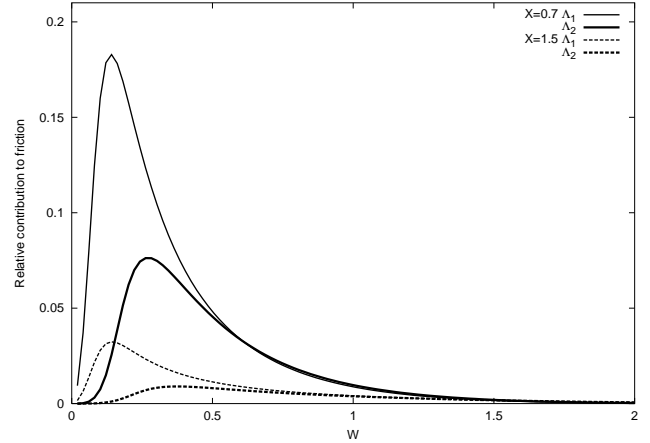}}
  }
\caption[]{
The function $\ln(\Lambda)g_{\rm ort}(X,W)$ for the orthogonal component of the 
inhomogeneous dynamical friction as a
function of encounter velocity $W$ for different satellite velocities
$X=0.7,1.5$ is shown. The difference from using $\Lambda_1$ (thin lines) or
 $\Lambda_2$ (thick lines)
is large and depends also on $X$ and on the system parameters
$q_{\rm d}$, $q_{\rm s}$.
 The same values as in
Fig. \ref{justfiggcha} are used.
}
\label{justfiggort2}
\end{figure}
For the inhomogeneous force the effect of the Coulomb logarithm on both
components ($g_{\rm par}(X,W)$ and $g_{\rm ort}(X,W)$) is similar. In Fig.
\ref{justfiggort1} the parameter dependence of the Coulomb logarithm on the
cut-off is demonstrated. The results on the contributions to the orthogonal
component are shown in Fig.
\ref{justfiggort2} using the
same parameters as in Fig. \ref{justfiggcha}. The main effect of a varying 
Coulomb
logarithm (with $Q_1$ or $Q_2$) is a well defined lower cut-off consistent with
the determination of $g_{\rm ort}(X,W)$ itself. The different approximations give
significantly differing results. Therefore we will discuss both cases in detail.
Using the approximation of a velocity independent Coulomb
logarithm $\ln\Lambda_0$ can only be used together with a somewhat arbitrary
cuttoff. This is not appropriate for applications due to the strong sensitivity
of the resultant force on the choice of the cut-off value (see below).

\begin{figure}[t]
\centerline{
  \resizebox{0.98\hsize}{!}{\includegraphics[angle=270]{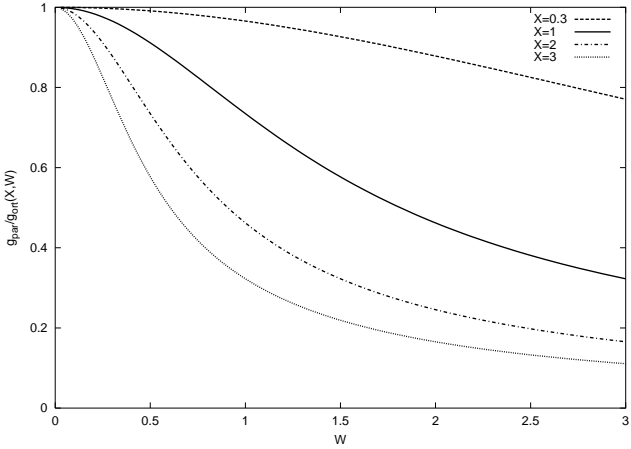}}
  }
\caption[]{
Here the ratio $g_{\rm par}(X,W)/g_{\rm ort}(X,W)$ of the contributions to the parallel
 and orthogonal component of
the inhomogeneous dynamical friction as a
function of encounter velocity $W$ for different satellite velocities
$X=0.3,1.0,2.0,3.0$ is shown. 
}
\label{justfiggorat}
\end{figure}
In Fig. \ref{justfiggorat} we show the contribution to the parallel force as a
function of $W$
relative to that for the orthogonal component. 
The parallel force is smaller for higher satellite
velocities $X$, because the contribution from higher encounter velocities $W$ 
is reduced with increasing $X$.
Since the Coulomb logarithm determines mainly the contribution from small $W$,
the effect on the orthogonal and the parallel components are similar .

\subsection*{Integration over $V_{\rm 0}$}

We introduce now also the integrals of the normalized functions defined in Eqs.
\ref{gcha}, \ref{gpar}, \ref{gort} by
\bq
G_{\rm hom,par,ort}(X) = \int_0^\infty  \ln (\Lambda )
g_{\rm hom,par,ort}(X,W) \dd W \label{Gfun} 
\eq
to get a convenient form for a discussion of the structure of the different
contributions to dynamical friction (see Eqs. \ref{vcha}, \ref{vpar},
\ref{vort}). With a velocity-dependent Coulomb logarithm, the integration over
$W$ must be done numerically. Here we give the derivation of the standard
friction formula and a short discussion of the behaviour of the inhomogeneous
terms at low values of $W$. The explicit form of the integrals for the
different contributions are given in Sect. \ref{dyfric} (Eqs. \ref{Ghom},
\ref{Gpar}, and \ref{Gort}). Further
discussion of their properties is also given there.

\subsection*{Chandrasekhar's friction formula}

With the approximation of a Coulomb logarithm
$\ln\Lambda_0$
independent of $W$ (Eq. \ref{l0}) we find the
standard Chandrasekhar friction formula by integrating $g_{\rm hom}(X,W)$ (Eq. 
\ref{gcha}) by parts
\bqn
&&\int_0^{\infty} g_{\rm hom}(X,W) \dd W =
        \frac{2}{\sqrt{\pi}}\int_0^{\infty}\exp(-W^2-X^2)\times
        \nonumber\\
&&\qquad        \left(\frac{\cosh(2WX)}{WX} -
                \frac{\sinh(2WX)}{2W^2X^2}\right)\dd W \nonumber \\
                &&= \frac{2}{\sqrt{\pi}X} \left[
                \left.\exp(-W^2-X^2)\frac{\sinh(2WX)}{2WX}\right|_0^{\infty}
                \right. \nonumber \\&& \qquad \left.
                +\frac{1}{X}\int_0^{\infty}\exp(-W^2-X^2)\sinh(2WX)\dd W \right]
                \nonumber \\
        &&= \frac{2}{\sqrt{\pi}X} \left[ -\exp(-X^2)+\frac{1}{2X}
        \int_{\rm -X}^{X}\exp(-Y^2)\dd Y \right]
                \nonumber \\
                &&= \frac{1}{X^2} \left[ erf(X) 
                -\frac{2X}{\sqrt{\pi}}\exp(-X^2)\right] \,.
\eqn
Inserting this into Eq. \ref{vch1} we find the result (Eq. \ref{Cha0})
\bq
\vec{\dot{v}_{\rm hom}} \approx
-C_{\rm Ch} \ln \Lambda_0 \frac{ \vec{e_{\rm v_{\rm M}}}}{X^2}
   \left[erf(X) - \frac{2X}{\sqrt{\pi}}\exp(-X^2)\right] \,.   
\eq
Since the maximum impact parameter depends on the local scale-length, the
Coulomb logarithm depends on the position of the satellite. Neglecting this
dependence also by using the global value $\langle\ln\Lambda\rangle$ leads to
the standard formula.

\subsection*{Inhomogenous force and the lower cut-off}

The functions $g_{\rm par}(X,W)$ and $g_{\rm ort}(X,W)$ show a singularity $\propto
W^{-2}$ for vanishing $W$ (Eqs. \ref{gpar}, \ref{gort}). Since the velocity-dependent Coulomb logarithms with
$Q_1$ and $Q_2$ vansih for small $W$, i.e. $\ln\Lambda_1\propto W^4$ and 
$\ln\Lambda_2\propto W^6$, there is a natural effective lower cut-off for the
integration over $W$. But the main contribution arises from low $W$ just above
this cut-off. This leads to differences of a factor of two at $X=0$, when using
the corresponding best fit values $Q_1=0.9$ and $Q_2=2.5$ for the orbit.

If we want to use a velocity-independent Coulomb logarithm $\ln \Lambda_0$, we
must introduce a lower cut-off $W_0$. With a similar approximation as used by
Binney (\cite{bin77} Eq. (7)), we find for $X\ll 1$ the corresponding formula
\bq
\dot{v}_{\rm inh}(X)\approx C_{\rm Ch}\frac{8}{3\sqrt{\pi}}\frac{1}{q_{\rm d}q_{\rm s}}
\frac{\langle\ln \Lambda\rangle}{W_0}\exp(-X^2) \,.
\eq
Inserting the parameters at pericentre of our standard orbit $q_{\rm d}=6$,
$q_{\rm s}=5$, $\langle\ln \Lambda\rangle=2.1$, and $G_{\rm ort}(X=0)=0.5$, we find the cut-off value
 $W_0=0.21$. We may compare this value with the relative velocity $W_h$ for a
 90-degree deflection at the half-mass radius $r_h$, which is
 \bq
 W_h=\sqrt{\frac{GM}{2\sigma^2 r_h}}= \sqrt{\frac{2}{q_{\rm s}}} =0.63
 \eq
 for the parameters above. Using $W_h$ would decrease the inhomogeneous friction
 force by a factor of three.

\end{document}